\documentclass[sigplan,10pt]{acmart}
\acmSubmissionID{<PAPER ID>}

\usepackage[english]{babel}
\usepackage{lipsum}    

\usepackage[normalem]{ulem}
\usepackage[capitalise]{cleveref}

\usepackage{tikz}
\usetikzlibrary{shapes,decorations.text,patterns}
\usepackage{adjustbox}

\usepackage{caption}

\usepackage{xspace}
\usepackage{subcaption}

\usepackage{tabularx}
\usepackage{multirow}
\usepackage{longtable}

\usepackage{algorithm}
\usepackage{algpseudocode}

\usepackage{amsmath}
\usepackage{centernot}

\usepackage[capitalise]{cleveref}
\usepackage{epigraph}
\usepackage{listings}
\usepackage[]{hyperref}

\definecolor{mygreen}{rgb}{0,0.6,0}
\definecolor{mygray}{rgb}{0.5,0.5,0.5}
\definecolor{mymauve}{rgb}{0.58,0,0.82}

\crefname{section}{\S}{\S\S}

\renewcommand\footnotetextcopyrightpermission[1]{} 
\newcommand{\showCODEN}[1]{\unskip}
\newcommand{\showDOI}[1]{\unskip}
\newcommand{\showLCCN}[1]{\unskip}
\newcommand{\showURL}[1]{\unskip}

\providecommand{\etal}{\emph{et al. }} 
\def\spacehack{0}
\ifnum\spacehack=1
\newcommand{\fillgap}[1]{{\vspace{#1}}}
\else
\newcommand{\fillgap}[1]{}
\fi

\def\shownotes{1}

\ifnum\shownotes=1
\newcommand{\authnote}[2]{{ $\ll$\textsf{\footnotesize #1 notes: #2}$\gg$}}
\else
\newcommand{\authnote}[2]{}
\fi

\newcommand{\codesm}[1]{\texttt{\small #1}}

\newcommand{\mnote}[1]{{\color{red}{\bf{\authnote{Mike}{#1}}}}}
\newcommand{\anote}[1]{{\color{blue}{\bf{\authnote{Anjali}{#1}}}}}
\newcommand{\toolname}{LockScope\xspace}

\providecommand{\host}{\textit{host}}
\providecommand{\runc}{\textit{runc}}
\providecommand{\runsc}{\textit{runsc}}

\providecommand{\vs}{vs. }
\providecommand{\ie}{\emph{i.e.,} }
\providecommand{\eg}{\emph{e.g.,} }
\providecommand{\etc}{\emph{etc. }}

\providecommand{\myparab}[1]{\vspace{0.2cm}\noindent\textbf{#1}}

\newenvironment{packeditemize}{\begin{list}{$\bullet$}{\setlength{\itemsep}{0.2pt}\addtolength{\labelwidth}{0pt}\setlength{\leftmargin}{\labelwidth}\setlength{\listparindent}{\parindent}\setlength{\parsep}{1pt}\setlength{\topsep}{0pt}}}{\end{list}}

\begin{document}
\widowpenalty=5000
\clubpenalty=5000

\title{\textsf{Locked In, Leaked Out: Measuring Isolation via Kernel Locks}}

\author{Anjali}\affiliation{University of Wisconsin-Madison}\email{anjali@wisc.edu}
\author{Michael M. Swift}\affiliation{University of Wisconsin-Madison}\email{swift@cs.wisc.edu}

\begin{abstract}
Isolation is a critical property for shared infrastructure to limit exposure and interference among simultaneous running workloads. Cloud providers use different isolation mechanisms such as full Virtual Machines, microVMs, Linux containers, secure containers \etc to confine workloads running 
in a multi-tenant environment. 

We propose a novel way to understand and measure performance interference and isolation at the system software layer that occurs due to shared access to data structures. We observe that interference takes place through shared structures, 
such as a kernel-level data structure, and that operating systems must synchronize access to these structures for safety. By measuring the level of synchronization between 
workloads, we can measure their ability to interfere and thus the amount of isolation the platform provides.

We demonstrate our method for measuring isolation by measuring the accesses to locks acquired in common across multiple workloads which indicates the amount of sharing through kernel data structures and hence the interference/isolation between two workloads. Furthermore, we identify the isolation properties of different kernel structures under different workloads, and find that the file system journal and kernel page allocator are the most common sources of interference.


\end{abstract}

\maketitle

\section{Introduction}
\label{sec:intro}

Isolation is essential in cloud environments where mutually distrusting tenants share physical hardware. Cloud providers use various isolation mechanisms to provide isolation between co-located workloads while maintaining performance guarantees. However, tightening the isolation boundary by running a workload in a virtual machine (VM) compared to running in a container can lower the overall resource utilization of a host. It is difficult to gather information about the amount of isolation a workload needs to minimize sharing with different co-located workloads. When sharing is less, there is less interference; hence, workloads are more isolated. 

 
Lightweight isolation platforms like Linux Containers, gVisor, and Firecracker rely on the host operating system for various resources and functionalities. This dependency leads to sharing and the potential for interference, which may have a detrimental impact on workloads by making them more prone to security attacks and performance degradation. Therefore, to reduce interference, it is important to understand what is being shared at the shared platform layer, \ie host kernel.


Two common sources of interference through OS resources are: (1) resource allocation, and (2) shared access to objects. The former arises when a workload needs a resource but is not able to due to unavailability as other co-running workloads use the resource. This type of interference is usually solved by leveraging OS (\eg cgroups, qdisc) and hardware isolation mechanisms (\eg cache partitioning) to partition resources to eliminate/reduce interference~\cite{ref:parties, ref:hcloud, ref:bolt, ref:untangle, ref:ubik, ref:vantage}. 
Resource-partitioning does not eliminate interference due to shared access to an object. Interference via access to objects happens when concurrent workloads want to access the same shared resource, such as a data structure within the host kernel. 

Our work tackles interference from shared accesses to data structures.  
We choose to analyze interference at the software stack because configuring underlying hardware for sharing and isolation (\eg controlling the DRAM bandwidth or applying cache-partitioning schemes) is mostly the same across platforms. The software stack is where the implementation of these platforms differ, \eg a userspace kernel in gVisor, and these software implementation choices impact how shared accesses vary across platforms. Also, we believe much research has already looked at sharing and isolation at the hardware layer~\cite{ref:utility_cache, ref:heracles, ref:untangle}. We are unaware of any work that has studied interference through shared access to objects. We note that past research~\cite {ref:fxmark, ref:commutativity_rule_mit} has analyzed kernel lock accesses for scalability. Our focus in this work is on studying interference through kernel locks. 

We show in Figure~\ref{fig:filebench_perf} the importance of 
interference via shared access to objects across different isolation platforms by stressing the filesystem for metadata operations
~\cite{ref:container_isolation_blog}. We run conflicting workloads for some periods and observe that such interference can lead to performance degradation. This type of interference can also lead to security vulnerability, leading to framing attacks~\cite{ref:tratr} or denial of service.



In this work, we do a comparative study to understand sharing and interference observed at the system software level via shared objects when concurrent workloads execute in different isolation platforms. 
Furthermore, we use the data on sharing and interference to understand isolation among concurrent workloads via shared kernel data structures. Our focus is on the {\em isolation of system structure}, meaning how well isolated the data structures of system software executing on behalf of a workload are. As many hardware isolation issues are orthogonal to the software structure (\eg micro-architectural side-channels), we do not measure them in this work~\cite{ref:sice, ref:realms, ref:avf-1, ref:avf-2}.

The first challenge in analyzing isolation at the system software layer is to recognize what resources/data structures/objects are shared among concurrent workloads.
We observe that limited isolation implies that workloads can interfere, so we look at the opportunities for and frequency of interference between workloads on a platform.  For example, two workloads accessing a shared cache in the OS can affect each other's behavior by modifying the cache contents. The key insight of our work is that most forms of interference take place on some shared data, and that access to shared data requires synchronization. As a result, we can identify points of interference by the prevalence of synchronization in workloads. Workloads that rarely synchronize share little data, have little impact on each other, and are well isolated. In contrast,  workloads that synchronize frequently (\eg locking access to shared data) have more opportunity to interfere by changing shared data or stalling waiting for locks; hence they are less isolated. One side effect of this approach is that it recognizes that isolation is workload dependent: two workloads executing purely user-mode code without making system calls may be perfectly isolated in operating system processes, while workloads that make heavy use of system services and multiple processes, accessing more shared data, benefit from more powerful isolation mechanisms. 

We make the following contributions in this work:
\begin{packeditemize}
\item{We collect and analyze kernel-level lock traces to measure system-level sharing via objects across a diverse set of applications and isolation platforms.}
\item{We identify memory allocation and file system journaling as the dominant sources of cross-application interference.}
\item{We implement a tool that enables fine-grained detection of software-level interference via dynamic kernel tracing.}
\end{packeditemize}


\section{Isolation and Interference Through Sharing}
\label{sec:motivation}

Isolation in computer systems keeps workloads apart from each other, so that the behavior of one workload does not affect other workloads. 
The goal for perfect isolation is {\em
non-interference}~\cite{ref:noninterference, ref:difc}, meaning that one workload has no impact on what another workload experiences. While often used for security 
by looking at what is possible for an attacker, for isolation we look more practically at the performance impact of one specific workload on another specific workload. Thus, an 
imperfect isolation mechanism can provide perfect non-interference if the two workloads do not make use of facilities that allow interference.

\begin{table}[]
\small
\begin{tabular}{|l|l|l|l|l|}
\hline
\begin{tabular}[c]{@{}l@{}} \textbf{Platform /}\\ \textbf{Resource}\end{tabular} & \textbf{KVM}                                                     & \textbf{runc}                                                       & \textbf{gVisor}                                                     & \textbf{Firecracker}                                             \\ \hline
\textit{Kernel}                                                         & Isolated                                                & Shared                                                     & Shared                                                     & Isolated                                                \\ \hline
\textit{Network}                                                        & \begin{tabular}[c]{@{}l@{}}Same as \\ host\end{tabular} & \begin{tabular}[c]{@{}l@{}}Same as \\ host\end{tabular}    & \begin{tabular}[c]{@{}l@{}}Shared/\\ Isolated\end{tabular} & \begin{tabular}[c]{@{}l@{}}Same as \\ host\end{tabular} \\ \hline
\begin{tabular}[c]{@{}l@{}}\textit{Memory} \\ \textit{Allocation}\end{tabular}   & Isolated                                                & Shared                                                     & \begin{tabular}[c]{@{}l@{}}Shared/\\ Isolated\end{tabular} & Isolated                                                \\ \hline
\textit{Filesystem}                                                     & Isolated                                                & \begin{tabular}[c]{@{}l@{}}Shared/\\ Isolated\end{tabular} & \begin{tabular}[c]{@{}l@{}}Shared/\\ Isolated\end{tabular} & Isolated                                                \\ \hline
\textit{Hardware}                                                       & Shared                                                  & Shared                                                     & Shared                                                     & Shared                                                  \\ \hline
\end{tabular}
\caption{Sharing across different platforms.} 
{\vspace{-0.3in}}
\label{tab:sharing_level}
\end{table}

The basic approach that isolation systems take is to separate workloads in space and time. Spatial isolation ensures workloads access different data, while temporal isolation prevents them from accessing the same data simultaneously.
Much as an operating system process provides separate address spaces and collections of resources, 
stronger forms of isolation provide increasingly more private resources and fewer shared resources. At the extreme, running workload on separate physical machines provides 
100\% private resources and no shared resources. This approach also dictates that resource efficiency is usually opposed to isolation: as a platform increases isolation, 
it decreases the amount of sharing across workloads that provides the opportunity for interference. Thus, increasing isolation decreases resource efficiency, which provides 
a strong motivation to find the {\em right level of isolation} for a workload --- using a too-strong platform wastes resources. On the other hand, if you increase efficiency by sharing more, you are increasing your risks for performance and security isolation. Table~\ref{tab:sharing_level} shows the resources that are shared/isolated across different isolation mechanisms. 

To reduce interference cloud providers use various isolation mechanisms such as full VMs, containers, microVMs \etc All this mechanism takes the approach of reducing the interactions with the shared platform \ie the host kernel across workloads to minimize interference and sharing. 

Building on this insight, we propose to measure the isolation of workloads on a platform by looking at the level of synchronization between the workloads. If there is little 
synchronization, then there is little access to shared data, and the workloads are well isolated. In contrast, if there is frequent, fine-grained synchronization, then there is poor 
isolation, as there are many opportunities. Thus, by analyzing the usage of locks in common across workloads in an isolation platform, we can measure the amount of isolation offered 
by the platform. We can also quantitatively compare the isolation offered by different platforms by looking at differences in synchronization behavior between platforms.

\section{Synchronization as an Identifier of Sharing}
\label{sec:locks}

\begin{table*}[t]
  \small
\begin{tabular}{p{4.3cm}p{5.3cm}p{6.2cm}} 
\begin{lstlisting}[caption={Private}]
void set_fs_pwd(
 struct fs_struct *fs,
 struct path *path){
 struct path old_pwd;
 path_get(path);
 spin_lock(&fs->lock);
 // sets fs->pwd to *path 
 spin_unlock(&fs->lock);
}

\end{lstlisting}&
\begin{lstlisting}[caption={Shared}]
void release_task(
struct task_struct *p){
 write_lock(&tasklist_lock);
 // removes task from list 
 write_unlock(&tasklist_lock);
}
\end{lstlisting}&
\begin{lstlisting}[caption={Incidentally Shared}]
void insert_inode_hash(
 struct inode *i, u64 hval){
 struct hlist_head *b =
 inode_hashtbl + hash(i->i_sb,hval);
 spin_lock(&inode_hash_lock);
 hlst_add_head_rcu(&i->i_hash,b);
 spin_unlock(&inode_hash_lock);
}
\end{lstlisting}
   \vspace{-1.4cm}
  \end{tabular} 
    \caption{Example lock usage in the Linux kernel.}
    {\vspace{-0.25in}}
    \label{tab:lock_acceses}
\end{table*}

\myparab{Interference and synchronization.} 
Within a software platform, interference implies that some shared resource between the two workloads 
acts as the agent for interference. Our key insight is that operating systems 
{\em synchronize all access to shared resources}, whether a physical
resource (memory, devices) or a logical resource (an object). Thus, any interference between 
workloads will be controlled by operating system synchronization.
Some examples include two processes sharing memory must synchronize around the physical pages being shared. Likewise, two processes 
sharing access to a device, such as a sound or a network card, must synchronize their access to the device. And, two processes sharing access to a file must synchronize access to the file system. 
Read sharing does not lead to interference, as each process can logically operate on a copy of data (\eg in a local processor cache) without any effect from other processes.

Furthermore, workloads that only incidentally 
access shared operating system data structures must still synchronize. 
Two programs accessing a shared file system, even if they access 
different files, may synchronize access to shared caches such as the Linux \codesm{dcache} or \codesm{inode} cache. 

\myparab{Locks and Synchronization.}
We note that shared resources in the kernel need synchronization to ensure a safe and correct order of execution of concurrent processes in the system. There are various primitives implemented in the Linux kernel to achieve synchronization:
\begin{packeditemize}
    \item Lock-based synchronization: These include spinlocks (and variants such as reader-writer and seqlocks), mutexes, and semaphores. 
    \item Lockless synchronization: These include atomic operations, Read-Copy Update (RCU), and memory barriers. 
\end{packeditemize}

Given the non-blocking nature of lockless mechanisms~\cite{ref:lockless_1, ref:lockless_2, ref:non_blocking_1}, they lead to less interference and contention compared to lock-based approaches. Locks enforce mutual exclusion and hence more contention, so their impact on interference and isolation will be more. 

Table~\ref{tab:lock_acceses} shows examples (code snippets) of private, shared, and
incidentally shared locks in the Linux kernel. The
\codesm{set\_fs\_pwd} function acquires a {\em private} lock, \codesm{fs->lock}, to protect the \codesm{fs->pwd} field of the
\codesm{fs\_struct} data structure that maintains the file system state for a process. This lock is not accessed by other processes. 
\codesm{tasklist\_lock}
is {\em shared} and synchronizes access to a resource shared by multiple processes, the list
of all tasks. In the example code, it is used to remove a task from
the list when the process exits. Finally, processes may {\em
  incidentally} synchronize by hashing into the same hash bucket, a form of false sharing,
although the protected data is not shared. \codesm{insert\_inode\_hash} acquires
 \codesm{inode\_hash\_lock} that protects the shared \codesm{inode\_hashtable}. 
 
We use lock-based synchronization as a proxy for sharing. While this will not cover all types of sharing, we believe that our work is a first step in understanding and quantifying sharing via synchronization, and can be extended to other synchronization primitives in the future. 

\section{Lock Tracing Implementation}
\label{sec:implementation}

\begin{figure}
  \includegraphics[width=0.5\textwidth]{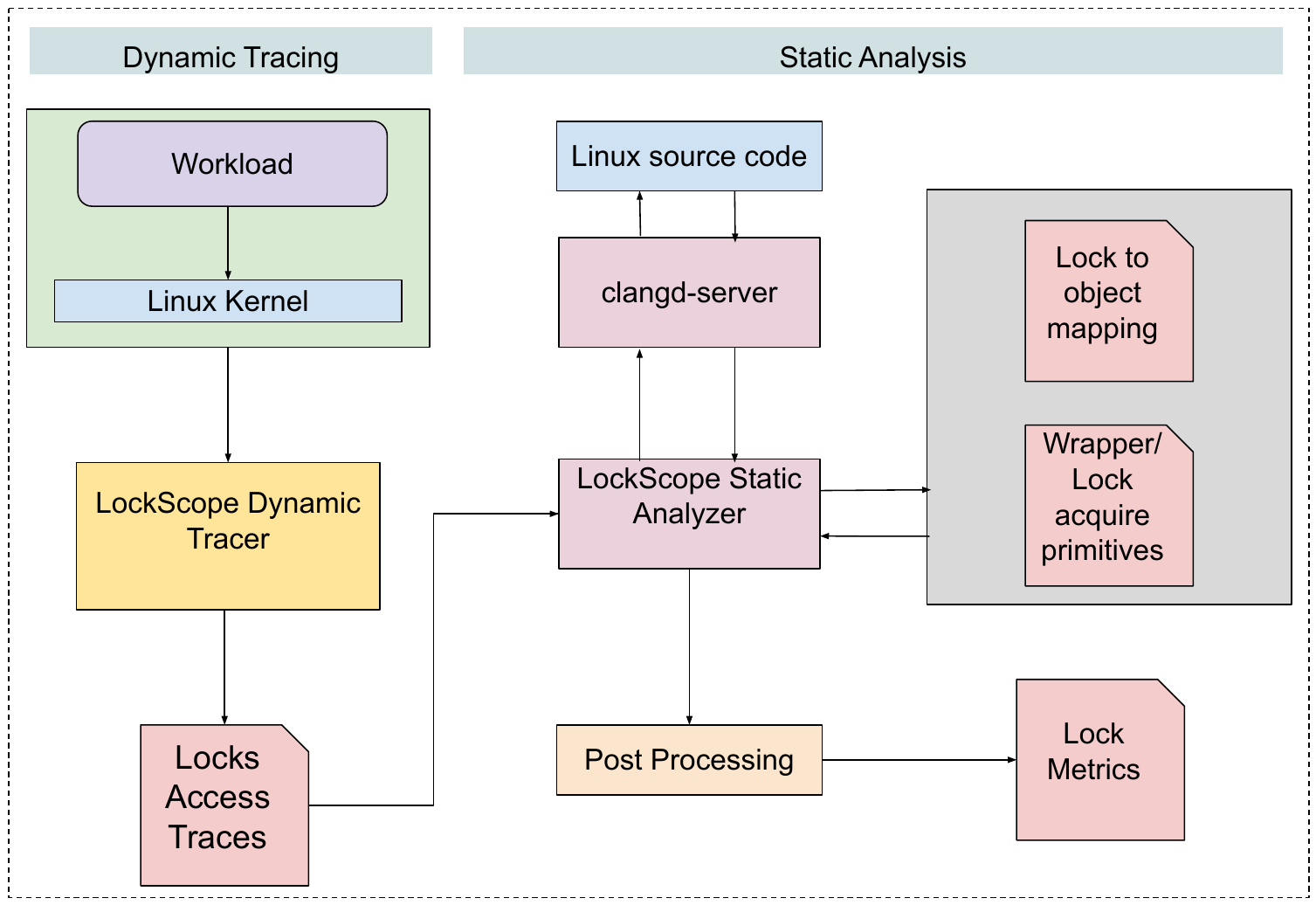}
  {\vspace{-0.2in}}
  \caption{Overview of {\em \toolname} workflow. In the dynamic tracing phase, \toolname collects lock accesses across workloads, and the static analysis phase completes the lock-to-object mapping.}
  {\vspace{-0.13in}}
  \label{fig:tracing-overview}
\end{figure}


\begin{table*}[]
\small
\begin{tabular}{|l|l|}
\hline
\textbf{Lock Type} & \textbf{Lock Primitives}                

\\ \hline
Spinlock           & \begin{tabular}[c]{@{}l@{}}\_raw\_spin\_lock, \_raw\_spin\_lock\_irqsave, \_raw\_spin\_lock\_irq, \_raw\_spin\_lock\_bh, \_raw\_spin\_trylock, \\ \_raw\_spin\_trylock\_bh,\_raw\_spin\_lock\_nested, \_raw\_spin\_lock\_irqsave\_nested,\end{tabular}                                                                \\ \hline
Read-write lock    & \begin{tabular}[c]{@{}l@{}}\_raw\_read\_lock, \_raw\_write\_lock, \_raw\_read\_lock\_bh, \_raw\_write\_lock\_bh, \_raw\_read\_lock\_irq, \\ \_raw\_write\_lock\_irq, \_raw\_read\_lock\_irqsave, \_raw\_write\_lock\_irqsave, \_raw\_read\_trylock, \\ \_raw\_write\_trylock, \_raw\_write\_lock\_nested\end{tabular} \\ \hline
Mutex              & \begin{tabular}[c]{@{}l@{}}mutex\_lock\_nested/mutex\_lock, rt\_mutex\_lock\_nested/rt\_mutex\_lock, mutex\_trylock, \\ rt\_mutex\_trylock, mutex\_lock\_interruptible\_nested\end{tabular}                                                                                                                           \\ \hline
Semaphore          & \begin{tabular}[c]{@{}l@{}}down\_read,  down\_write, down\_read\_trylock, down\_write\_trylock, down\_read\_nested, \\ down\_write\_nested, down\_read\_killable, down\_write\_killable, down\_read\_killable\_nested,\\ down\_write\_killable\_nested, down\_read\_interruptible\end{tabular}                        \\ \hline
\end{tabular}
\caption{Lock primitives probed by {\em \toolname} dynamic tracer.}
\label{tab:lock_primitives}
 {\vspace{-0.5cm}}
\end{table*}

\begin{table}[]
    \small
\begin{tabular}{|l|l|l|l|}
\hline
file     & object\_name & lock\_type                 & lock\_name
\\ \hline
fs/ext4/ext4.h & ext4\_inode\_info & rw\_semaphore& i\_data\_sem       \\ \hline
\end{tabular}
\caption{An example row in the lock to object mapping generated using symbol information.}
{\vspace{-0.4in}}
\label{tab:example_struct_lock_map}
\end{table}

We implemented a data-collection tool named {\em \toolname} to measure the lock usage of workloads running on isolation platforms on Linux. The goal of \toolname is to trace lock acquisitions dynamically and use static analysis to identify location in code where a lock acquisition in the trace occurs and, more importantly, what objects the lock protects. 
We show an overview of {\em \toolname} in Figure~\ref{fig:tracing-overview}. {\em \toolname}  consists of two components: (a) dynamic tracer and (b) static analyzer. {\em \toolname} captures and analyzes fine and coarse-grained runtime locking activity, enabling the precise identification of object-level interference across platforms.

\subsection{\toolname Dynamic Tracer}
We capture the dynamic lock usage using a modified and extended version of {\em klockstat}, an eBPF lock monitoring tool~\cite{ref:klockstat}. This tool traces lock acquire and release functions in the Linux kernel and can be attached to mutexes, RT-mutexes, semaphores, spinlocks, and reader-writer spinlocks. Table~\ref{tab:lock_primitives} lists all the lock primitives we probe with klockstat.

The {\em \toolname} dynamic tracer produces a trace of every lock acquired, and for each 
operation, records: the process and thread IDs, lock address, lock name, kernel stack trace, acquire time, hold time, and count of occurrences of that
stack trace with that lock. In addition, we instrument lock
initialization routines to learn when a lock address is
reallocated for a new lock. We process this trace to calculate different points using interference metrics such as the 
number of {\em shared locks} --- locks held in common
across workloads --- accessed for different workloads and {\em private locks} --- unique and non-shared locks held
by each workload. 

In summary, dynamic analysis produces a trace of all the locks acquired by a running workload. Below is an example of the trace output. 
The stack details are stored in a different file and matched with the stack\_id to get the stack traces with an acquisition. 
We find the shared locks using the unique lock addresses common to workloads.
\begin{small}
\begin{tabular}{|l|l|l|l|l|l|}
\hline
PID   & addr            & name                 &  count & process & stack \\ \hline
38296 & 0xffff888 & \&pipe-\textgreater{}mutex &  1           & mmap04  & 324596    \\ \hline
\end{tabular}
\end{small}

\subsection{\toolname Static Analyzer}
The {\em \toolname} static analyzer resolves a stack trace from the dynamic tracer into the exact location in code where the lock was acquired and, when the locks is a member of an object, the object containing the lock. For global locks, we resolve the lock name and its definition.
We implemented the {\em \toolname} static analyzer using the
 clangd~\cite{ref:clangd} language server~\cite{ref:lsp}. clangd is widely used in many IDEs for different functionalities (\eg symbol indexing and outgoing call analysis), and it provides a convenient API for querying code properties. We use clangd for static analysis as it offers simple mechanism to accurately index into a large codebase, including macro resolution, type deduction, and cross-referenced navigation, providing functionality closer to a compiler-driven source code analysis compared to other available tools such as cscope\cite{ref:cscope}. 
 
The stack traces from the eBPF tools are sometimes incomplete \ie the trace stops 2-3 levels due to inline functions and certain optimizations such as tail call optimization~\cite{ref:ebpf-book, ref:ebpf_frame_pointer}. We use the static analysis tool to resolve the stack traces to the exact line numbers of lock acquisitions in the code. We use the lock names, file names, and function names to determine this. Furthermore, we use static analysis to identify the data structure a lock protects. We do this to learn about the nature of sharing of different data structures and also to understand which data is private \vs mostly shared.

{\em \toolname} static analyzer is a command-line utility tool and supports the following functionality: 
\begin{packeditemize}
\item \textit{Symbol Generation} produces all the details related to a symbol, including name, type, signature, and line number ranges within the source code where the symbol is defined. We use this to 
create a mapping of locks to the \texttt{struct} within which they are defined. 
Table~\ref{tab:example_struct_lock_map} provides an example row from this file. Moreover, we use the symbol information to generate a list of system-wide global locks ($\approx$2488).


\item \textit{Outgoing Calls} uses function-type symbol information to identify outgoing calls. We use this to resolve the point of acquisition for locks in cases of incomplete dynamic stacks, where the start function is the last function recorded in the collected stack.

\item \textit{Incoming Calls} works similar to \textit{Outgoing Calls}. We leverage this information to identify potential lock wrapper functions (335 identified so far), which serve as stop conditions when resolving the point of acquisition in incomplete dynamic stacks. 

\item \textit{AST} generates an abstract syntax tree for a source file, which {\em \toolname} uses to resolve lock to object mapping, where lock names result in many-to-many mapping in the file generated using document symbol (\eg lock name with \textit{lock} can be associated with multiple objects). We call these locks \textit{generic}. 

We use the AST of the function  where  \textit{generic} locks are acquired to get 
line number associated with the lock variable, which we use to query \textit{Get Definition} for the lock symbol to find the location of its definition. 

\item \textit{Get Definition} returns the file path and line number where the lock is defined. We use this to locate the corresponding object name by matching the line number to the range associated with a struct-type symbol from \textit{Symbol Generation}.

\end{packeditemize}

For all unique locks, identified by lock name, function name, and file, acquired across all workloads analyzed in this paper (574 in total), we can map approximately 75\% of them to their associated objects.
Our tool currently does not resolve cases where locks are acquired through macros or where object variables are reassigned to local variables within function bodies. We manually added support for a small number of such macros (about 5) that we encountered during our analysis.

\section{Locking Analysis Methodology}
\label{sec:method}
Our evaluation goals are:
\begin{packeditemize}
    \item What are the different locks and data structures accessed by a workload?
    \item What shared data is accessed most frequently? 
    \item How does sharing via objects vary across platforms for the same workload? 
    \item How can we quantify interference and isolation via shared accesses to objects using kernel locks?
\end{packeditemize}

\subsection{Platform and Workloads}
We run all our experiments on Cloudlab~\cite{ref:cloudlab} c220g5
machine, with twenty cores Intel Xeon Silver 4114 running at 2.20 GHz, 192GB ECC DDR4-2666 Memory, two disks: Intel DC  S3500 480 GB 6G SATA SSD and 1 TB 7200 RPM 6G SAS HDs, and 10Gbps NIC. We run on Ubuntu 22.04 (kernel v6.1). 

For each workload under test, we collect host kernel lock traces to understand how different isolation platforms use various kernel data structures for functionalities. We also run stress tests to analyze the impact of using shared kernel data structures on performance due to synchronization.

For all tests, we turn off SMT to minimize interference through resource contention and cross-socket communication. We also isolate the LLC by enabling Intel's Cache Allocation Technology (CAT). In our setup, we have 11 LLC way partitioning, and each Class of Service (COS) (1-11) is assigned to one cache line. All cores (on socket 0) have a one-to-one mapping with COS. Our focus is to understand interference caused due to shared access to kernel resources. Hence, our experiments are designed to measure interference via system structures and avoid hardware interference (\eg not exceeding local caches or disk bandwidth). 
We use the default kernel configuration for stress tests but enable lock-related debugging configurations (\eg CONFIG\_DEBUG\_SPINLOCK, CONFIG\_DEBUG\_MUTEXES) for collecting traces. 

We run workloads on four platforms: (a) host, (b) LXC (runc), (c) gVisor release-20231106.0 (runsc), and (d) Firecracker v1.10.1 (fc). For runsc, we use the KVM platform, which is recommended for bare-metal~\cite{ref:gvisor_platform}. Our goal is to understand object usage patterns across a diverse spectrum of system behavior (\eg stressing particular subsystems, light/medium/heavy kernel subsystem usage). To capture these diverse cases, we run our evaluation on three different sets of workloads across these platforms:

\myparab{Microbenchmarks.} We choose a set of microbenchmarks, stressing different kernel subsystems to understand object usage under targeted stress environments. 

\myparab{Serverless Workloads.} As serverless workloads are a major use case for lightweight isolation, we select applications (image and video processing), ML training (logistic regression), and ML serving (face detection, CNN, and RNN) workloads from FunctionBench suite~\cite{ref:funcbench} to analyze object usage across modern lightweight serverless functions. This is useful for gaining insights into short-lived object usage patterns in a FaaS environment.

\myparab{Cloud Workloads.} To capture object usage in heavy and long-running workloads, we run  cloud workloads as shown in Table~\ref{ref:cloud_workloads_desc}. Feedsim and VideoTranscode are taken from DCPerf~\cite{ref:dcperf}. We were not able to set up the analytics and data caching from DCPerf on gVisor due to its very restricted environment, so we replaced them with the Graph Analytics and Data Caching workloads from CloudSuite~\cite{ref:cloudsuite}.


\subsection{Lock Usage Tracing}
We trace workloads using the {\em \toolname} dynamic tracer and two simultaneous identical workloads on the same isolation platform. Each workload executes for a fixed duration. We re-execute the workloads once for each lock type described in Table~\ref{tab:lock_primitives} to get cleaner traces (\eg minimizing nesting) and better lock coverage. We collect the traces in three iterations to maximize the coverage of unique locks. We do not trace the platform startup.

We collect system-wide traces while the workload is running and then select entries belonging to the running workload. For firecracker, we select the traces belonging to the firecracker process. For gVisor, we use the sandbox and gofer pids to select the traces. For runc, we leverage the namespace filtering~\cite{ref:ebpf_ns} from eBPF to select the traces belonging to each container namespace and then further filter it to remove docker and runc-related management activities. Because locks acquired in interrupt context (\eg timer interrupts, softirqs) may not be caused by the current process, we remove these lock accesses.  We inspect the stack track of lock acquires and remove locks whose  traces include interrupt-related functions (\eg 
\_\_softirqentry\_text\_start,
hrtimer\_interrupt). 

We use the \toolname static analyzer to process the lock traces and complete the mapping of lock names to kernel objects. For resolving the point of acquisition for some incomplete dynamic traces, we use the outgoing calls functionality of the static analyzer and then do a breadth-first search from the last function in the stack traces. We use the lock primitives and the lock name to stop the search. We also leverage lock acquire function wrappers generated by the tool to stop the search in some cases.


We use lock traces to determine the following:
\begin{packeditemize}
    \item \textit{Shared and private locks:} We identify unique shared and private locks by their lock addresses and report the average count across runs.
    \item \textit{Locks access rate:}  We report lock access rates. For each lock (identified by name,  function, file, and type), we compute the rate by dividing the acquire count by execution time. We then sum these per-lock rates across runs to obtain a cumulative access rate, which helps quantify how frequently different locks are accessed.
    \item \textit{Lock access across kernel subsystems:} We group the lock rates by kernel subsystem (\eg \texttt{mm}, \texttt{fs}, \texttt{kernel}) based on file paths to understand how lock usage varies across different parts of the kernel.
\end{packeditemize}

\subsection{Performance Interference}
\label{eval:perf-interference-trasher}
We measure the performance of workloads with a {\em trasher} designed to stress kernel resources usage through frequent system calls. For each performance test (unless specified otherwise), we first start a workload (worker), and launch an additional trasher every ten minutes, causing increasing interference. For resources that allow interference, this leads to decreasing performance as trashers are added.
We pin the worker and trashers to different CPU cores sharing the same socket.

\section{Measuring Lock Usage}
\label{sec:locks_interference}
We begin by measuring the lock usage of microbenchmarks and application workloads to identify opportunities for interference. We run two copies of the same workload simultaneously and use \toolname to record the locks each copy acquires, and of those which are shared across the two instances and which are private to an instance. We show the average shared lock counts and the cumulative lock rate for all workloads traced in Table~\ref{tab:lock_count_rate}. For lock rate, we aggregate over all rates by calculating the sum.

\begin{table}[]
\small
\begin{tabular}{|r|rrrr|}
\hline
\multirow{2}{*}{\textbf{Workloads}} & \multicolumn{1}{r|}{\textbf{Host}} & \multicolumn{1}{r|}{\textbf{runc}} & \multicolumn{1}{r|}{\textbf{runsc}} & \textbf{fc} \\ \cline{2-5} 
 & \multicolumn{4}{c|}{\textbf{shared (rate)}} \\ \hline
mem-8KB & \multicolumn{1}{r|}{\begin{tabular}[c]{@{}r@{}}0.33 \\ (226)\end{tabular}} & \multicolumn{1}{r|}{\begin{tabular}[c]{@{}r@{}}0.33 \\ (0.0016)\end{tabular}} & \multicolumn{1}{r|}{\begin{tabular}[c]{@{}r@{}}1 \\ (73)\end{tabular}} & \begin{tabular}[c]{@{}r@{}}2.67 \\ (399)\end{tabular} \\ \hline
mem-1Mb & \multicolumn{1}{r|}{\begin{tabular}[c]{@{}r@{}}0.33 \\ (610)\end{tabular}} & \multicolumn{1}{r|}{\begin{tabular}[c]{@{}r@{}}2.33\\  (0.19)\end{tabular}} & \multicolumn{1}{r|}{\begin{tabular}[c]{@{}r@{}}1 \\ (220)\end{tabular}} & \begin{tabular}[c]{@{}r@{}}1.66\\ (121)\end{tabular} \\ \hline
file-list & \multicolumn{1}{r|}{\begin{tabular}[c]{@{}r@{}}2.66 \\ (321)\end{tabular}} & \multicolumn{1}{r|}{\begin{tabular}[c]{@{}r@{}}9.67 \\ (897)\end{tabular}} & \multicolumn{1}{r|}{} & \begin{tabular}[c]{@{}r@{}}1.66 \\ (1336)\end{tabular} \\ \hline
file-create & \multicolumn{1}{r|}{4 (29)} & \multicolumn{1}{r|}{\begin{tabular}[c]{@{}r@{}}15.67 \\ (1740)\end{tabular}} & \multicolumn{1}{r|}{} & \begin{tabular}[c]{@{}r@{}}1.67 \\ (2367)\end{tabular} \\ \hline
file-delete & \multicolumn{1}{r|}{3.66 (353)} & \multicolumn{1}{r|}{\begin{tabular}[c]{@{}r@{}}3.67 \\ (296)\end{tabular}} & \multicolumn{1}{r|}{} & \begin{tabular}[c]{@{}r@{}}1 \\ (3880)\end{tabular} \\ \hline
 & \multicolumn{1}{r|}{} & \multicolumn{1}{r|}{} & \multicolumn{1}{r|}{} &  \\ \hline
\begin{tabular}[c]{@{}r@{}}image\\  processing\end{tabular} & \multicolumn{1}{r|}{\begin{tabular}[c]{@{}r@{}}5.33 \\ (1799)\end{tabular}} & \multicolumn{1}{r|}{\begin{tabular}[c]{@{}r@{}}4.33 \\ (1012)\end{tabular}} & \multicolumn{1}{r|}{\begin{tabular}[c]{@{}r@{}}0.67 \\ (12)\end{tabular}} & \begin{tabular}[c]{@{}r@{}}0.67 \\ (3.751)\end{tabular} \\ \hline
\begin{tabular}[c]{@{}r@{}}video \\ processing\end{tabular} & \multicolumn{1}{r|}{\begin{tabular}[c]{@{}r@{}}7.33 \\ (3488)\end{tabular}} & \multicolumn{1}{r|}{\begin{tabular}[c]{@{}r@{}}1.99 \\ (1950)\end{tabular}} & \multicolumn{1}{r|}{\begin{tabular}[c]{@{}r@{}}0.67 \\ (0.14)\end{tabular}} & \begin{tabular}[c]{@{}r@{}}1.33 \\ (4.3)\end{tabular} \\ \hline
lr\_training & \multicolumn{1}{r|}{\begin{tabular}[c]{@{}r@{}}12.66 \\ (15)\end{tabular}} & \multicolumn{1}{r|}{\begin{tabular}[c]{@{}r@{}}2 \\ (17)\end{tabular}} & \multicolumn{1}{r|}{\begin{tabular}[c]{@{}r@{}}1.67 \\ (1.64)\end{tabular}} & \begin{tabular}[c]{@{}r@{}}0.66 \\ (0.86)\end{tabular} \\ \hline
face\_detection & \multicolumn{1}{r|}{\begin{tabular}[c]{@{}r@{}}6.67 \\ (80)\end{tabular}} & \multicolumn{1}{r|}{\begin{tabular}[c]{@{}r@{}}2.67\\  (12)\end{tabular}} & \multicolumn{1}{r|}{\begin{tabular}[c]{@{}r@{}}1.67 \\ (0.27)\end{tabular}} & \begin{tabular}[c]{@{}r@{}}0.33 \\ (0.01)\end{tabular} \\ \hline
cnn & \multicolumn{1}{r|}{\begin{tabular}[c]{@{}r@{}}3.33 \\ (4.06)\end{tabular}} & \multicolumn{1}{r|}{\begin{tabular}[c]{@{}r@{}}2.67 \\ (239)\end{tabular}} & \multicolumn{1}{r|}{\begin{tabular}[c]{@{}r@{}}1.67 \\ (13)\end{tabular}} & \begin{tabular}[c]{@{}r@{}}0.33 \\ (1.85)\end{tabular} \\ \hline
rnn & \multicolumn{1}{r|}{\begin{tabular}[c]{@{}r@{}}2.33 \\ (1.8)\end{tabular}} & \multicolumn{1}{r|}{\begin{tabular}[c]{@{}r@{}}1.67 \\ (2.82)\end{tabular}} & \multicolumn{1}{r|}{\begin{tabular}[c]{@{}r@{}}1.34\\  (0.59)\end{tabular}} & \begin{tabular}[c]{@{}r@{}}1 \\ (9.23)\end{tabular} \\ \hline
 & \multicolumn{1}{r|}{} & \multicolumn{1}{r|}{} & \multicolumn{1}{r|}{} &  \\ \hline
\begin{tabular}[c]{@{}r@{}}graph \\ analytics\end{tabular} & \multicolumn{1}{r|}{} & \multicolumn{1}{r|}{\begin{tabular}[c]{@{}r@{}}8.33 \\ (382)\end{tabular}} & \multicolumn{1}{r|}{\begin{tabular}[c]{@{}r@{}}5 \\ (419)\end{tabular}} & \begin{tabular}[c]{@{}r@{}}1.33 \\ (59)\end{tabular} \\ \hline
data caching & \multicolumn{1}{r|}{} & \multicolumn{1}{r|}{\begin{tabular}[c]{@{}r@{}}2.33\\  (0.008)\end{tabular}} & \multicolumn{1}{r|}{\begin{tabular}[c]{@{}r@{}}0.33 \\ (0.0011)\end{tabular}} & \begin{tabular}[c]{@{}r@{}}1.33 \\ (26)\end{tabular} \\ \hline
\begin{tabular}[c]{@{}r@{}}data caching \\ (no warmup)\end{tabular} & \multicolumn{1}{r|}{} & \multicolumn{1}{r|}{\begin{tabular}[c]{@{}r@{}}1.33 \\ (0.018)\end{tabular}} & \multicolumn{1}{r|}{\begin{tabular}[c]{@{}r@{}}0.33 \\ (0.0005)\end{tabular}} & \begin{tabular}[c]{@{}r@{}}1.33 \\ (1.41)\end{tabular} \\ \hline
feedsim & \multicolumn{1}{r|}{} & \multicolumn{1}{r|}{\begin{tabular}[c]{@{}r@{}}7.67 \\ (63)\end{tabular}} & \multicolumn{1}{r|}{\begin{tabular}[c]{@{}r@{}}3.33 \\ (43)\end{tabular}} & \begin{tabular}[c]{@{}r@{}}1\\  (0.62)\end{tabular} \\ \hline
\begin{tabular}[c]{@{}r@{}}video\\ transcode\end{tabular} & \multicolumn{1}{r|}{} & \multicolumn{1}{r|}{\begin{tabular}[c]{@{}r@{}}15 \\ (82)\end{tabular}} & \multicolumn{1}{r|}{\begin{tabular}[c]{@{}r@{}}8 \\ (3.06)\end{tabular}} & \begin{tabular}[c]{@{}r@{}}3.66 \\ (62)\end{tabular} \\ \hline
\end{tabular}
\caption{Shared average lock count and cumulative lock access rate (in parentheses) for each workload and platform.}
{\vspace{-0.25in}}
\label{tab:lock_count_rate}
\end{table}


\subsection{Microbenchmarks}
We use microbenchmarks targeting a particular OS resource to measure interference for those resources. For memory, we stress page allocation, and for files we stress metadata operations. We do not include CPU-bound workloads, as they typically make little use of OS resources and our experiments show little locking or interference. We do not measure lock usage in the networking subsystem, all our experiments are designed to minimize or eliminate networking access paths. 

\subsubsection{Memory}

To stress the memory subsystem, we use a memory microbenchmark that allocates a total of 16GB memory with different \texttt{mmap} allocation sizes. The microbenchmark touches one word of each allocated page by writing to it and finally, calls \texttt{munmap} to free the memory. The benchmark runs in a loop, continuously allocating and deallocating memory for a specified time duration. We allocate a total of 18GB to each running instance to avoid interference due to resource allocation and pin each workload to a separate core. We run the benchmark for two allocation sizes: 8K and 1MB. 

We observe that although shared lock counts for platforms remain low, two shared locks stand out. The global buddy allocation locks, \texttt{zone->lock}, is accessed by host, runsc, and fc at a high rate, the highest rate for host ($\approx$610 for 1MB), runsc ($\approx$220 for 1MB), and fc ($\approx$205 for 8KB). \texttt{zone->lock} protects a \texttt{zone} object, which holds the free list used for physical page allocations. This global lock can be the point of interference under high memory pressure, resulting in performance degradation for these platforms.  

Firecracker also acquires \texttt{lruvec->lru\_lock} with a high rate (about 193) for 8KB allocation size. The \texttt{lruvec} structure is used for maintaining pages in LRU order for a combination of memory zone and memory cgroups to select pages for reclamation. The high acquisition rate may be due to frequent updates to the LRU lists during the initial phases of page allocation, where newly allocated pages need to be inserted into the appropriate LRU list. This suggests that even early-stage memory usage in Firecracker involves significant interaction with reclamation-related data structures.

In contrast, runc exhibits minimal shared lock usage. This behavioral difference, compared particularly to the host arises because the host workload instance runs without memory limits, while the runc container is constrained via cgroups with an upper memory limit. The memory limit being pre-allocated in runc avoids triggering memory pressure and bypasses code paths involving the global allocator. 

\subsubsection{File Metadata Operations}

We measure file system metadata operations to understand object-level interference as they are more prone to performance degradation due to software-level interference ~\cite{ref:fxmark}.

We use Filebench~\cite{ref:filebench} to evaluate file metadata operations, and report the operations per second (ops) for each benchmark. Filebench requires address space layout randomization (ASLR) to be disabled. Currently, gVisor does not support disabling ASLR, so we were unable to get Filebench to run under gVisor and exclude it from this benchmark. Each workload instance operates on separate directories.

\myparab{List Directory.} This benchmark looks for contention around directory entries and inodes. It lists the contents of a large directory (50,000 entries). 

\myparab{Create Files.} This measures the performance of file creation operations by creating 50,000 files (4KB) in a directory. 

\myparab{Delete Files.} It uses a single thread to measure file deletion operation in a directory with 50,000 files

We observe a high rate of shared locks across operations, highest in runc, followed by host, and fc. We observe a high access rate for locks related to the filesystem bookkeeping and some memory management.  

The primary shared locks are associated with the file system  
\texttt{journal\_s} structure and  the superblock \texttt{ext4\_sb\_info}. While host and runc access both, fc only accesses the journal. These are single structures shared by an entire file system volume, so that even if workloads access different files, they will still access the same superblock and journal. 
\texttt{journal->j\_list\_lock}, which protects the per‑transaction list of modified buffers in the journaling layer, serializing access to the buffer, has the highest rate for the host ($\approx304$ for list).  \texttt{bgl->locks[i].lock} is a per‑block‑group spinlock acquired for any updates to a group’s block and inode bitmaps is another commonly accessed lock between host and runc. It has a lower access rate on both due to a fine-grained locking mechanism (acquired per block), but can still be a significant interference point under load. This result shows that fine-grained locking mechanisms are also prone to such interference, specifically under load, which can negatively impact performance. 

Uniquely, runc acquires \texttt{aa\_buffers\_lock} with a high rate (highest being $\approx$660 for create) for all operations that protect AppArmor's buffer pool used for security metadata, ensuring safe allocation and reuse of buffers. 
The global \texttt{inode\_hash\_lock}  accessed  by runc ($\approx$333 for create) and protects access to the global inode hash table. This result emphasizes that despite operating on different inodes, processes can still contend for this lock due to incidental sharing, highlighting how such shared structures can become significant points of interference. fc, on the other hand, mostly acquires \texttt{lruvec->lru\_lock} apart from  \texttt{journal->j\_list\_lock}, but the access rate is low.

Both host and runc acquire multiple shared locks due to their high reliance on a shared host filesystem, leading to potential interference and performance overhead during metadata operations. High sharing of locks such as \texttt{journal->j\_list\_lock} and \texttt{inode\_hash\_lock} can result in low performance isolation.
 
Firecracker has an isolated filesystem, which runs most operations inside the guest, and helps minimize lock usage. However, it still accesses some host kernel objects like \texttt{journal\_s} when writing back to the disk image, which creates potential channels for interference at the host level.

\subsection{Application Workloads}
The preceding results look at single-resource microbenchmarks. We now 
consider applications in two categories: serverless workloads
and some popular cloud workloads. 

\subsubsection{Serverless Workloads}

\begin{table*}[]
\small
\begin{tabular}{|l|l|lll|l|l|l|}
\hline
\multirow{2}{*}{\textbf{Category}} & \multirow{2}{*}{\textbf{Name}} & \multicolumn{3}{l|}{\textbf{Subsystem loads}} & \multirow{2}{*}{\textbf{Description}} & \multirow{2}{*}{\textbf{Input}} & \multirow{2}{*}{\textbf{Output}} \\ \cline{3-5}
 &  & \multicolumn{1}{l|}{\textbf{CPU}} & \multicolumn{1}{l|}{\textbf{Memory}} & \textbf{Disk} &  &  &  \\ \hline
\multirow{2}{*}{Application} & image processing & \multicolumn{1}{l|}{medium} & \multicolumn{1}{l|}{medium} & low & \begin{tabular}[c]{@{}l@{}}Image transformation using different \\ effects (PIllow)\end{tabular} & image & image \\ \cline{2-8} 
 & video processing & \multicolumn{1}{l|}{high} & \multicolumn{1}{l|}{high} & medium & Applies gray scale effect (OpenCV) & video & video \\ \hline
ML  training & logistic regression & \multicolumn{1}{l|}{high} & \multicolumn{1}{l|}{high} & medium & \begin{tabular}[c]{@{}l@{}}Review analysis and training (logistic \\regression, scikit-learn)\end{tabular} & text & model \\ \hline
\multirow{3}{*}{ML  Serving} & face detection & \multicolumn{1}{l|}{medium} & \multicolumn{1}{l|}{medium} & low & \begin{tabular}[c]{@{}l@{}}Annotates face in a video (CascadeClassifier, \\ OpenCV)\end{tabular} & video & video \\ \cline{2-8} 
 & cnn & \multicolumn{1}{l|}{medium} & \multicolumn{1}{l|}{medium} & low & \begin{tabular}[c]{@{}l@{}}Image classification (SqueezeNet, Tensorflow, \\ CNN)\end{tabular} & image & JSON \\ \cline{2-8} 
 & rnn & \multicolumn{1}{l|}{low} & \multicolumn{1}{l|}{low} & low & Words generation (PyTorch, RNN) & JSON & JSON \\ \hline
\end{tabular}
\caption{Serverless functions adopted from FunctionBench and vHive.}
{\vspace{-0.18in}}
\label{tab:funcbench}
\end{table*}

We run application and ML-related serverless workloads adopted from the FunctionBench suite~\cite{ref:funcbench} and vHive~\cite{ref:vhive}. We briefly describe each benchmark in Table~\ref{tab:funcbench}. We modified these workflows to read inputs and write output locally instead of S3. Hence, the network load 
is low, and disk I/O may be higher as we read/write data locally. 

\begin{figure*}[t]
\centering
     \begin{subfigure}[b]{0.24\textwidth}
         \centering
         \includegraphics[width=\columnwidth]{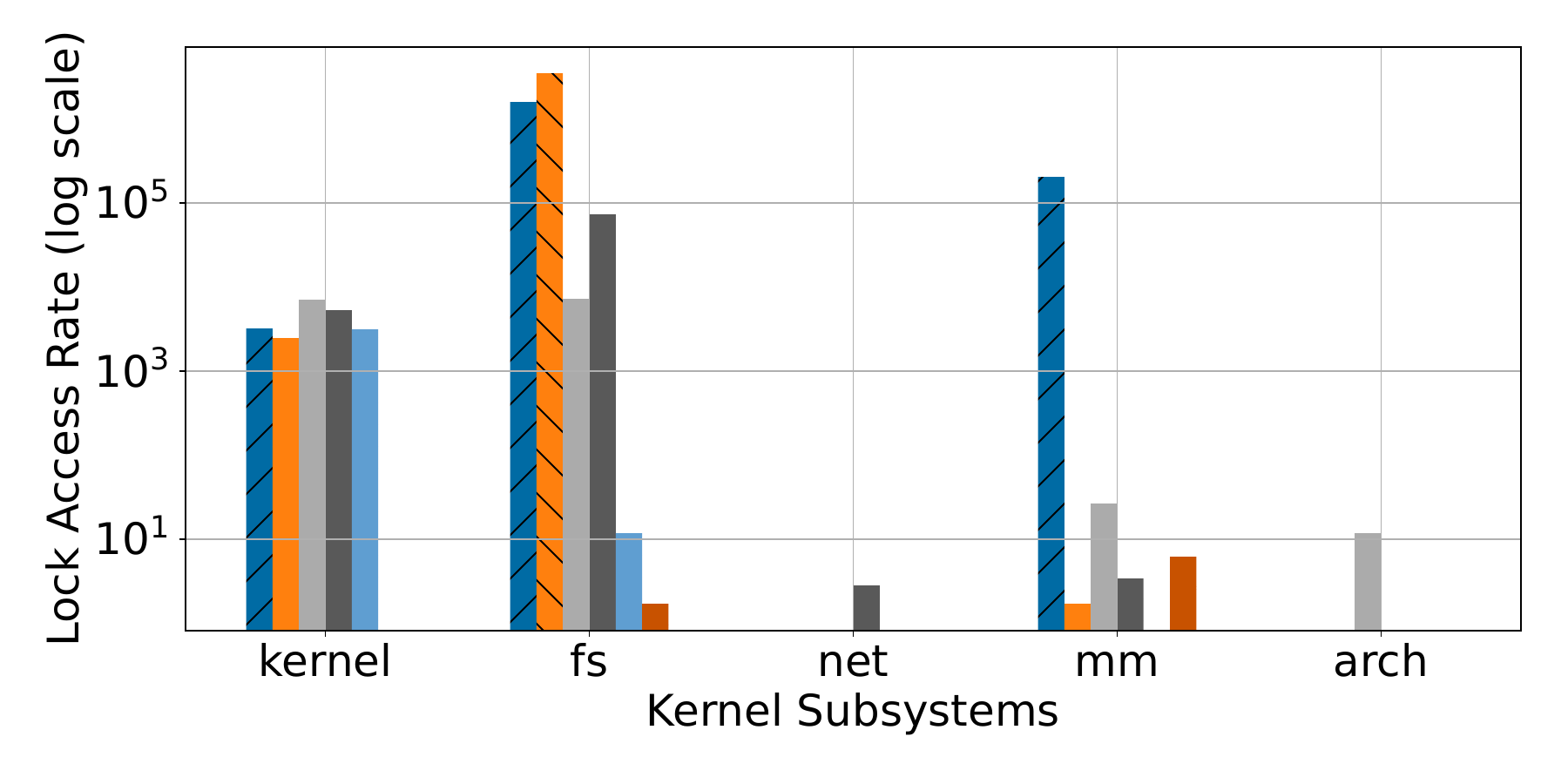}
          {\vspace{-0.25in}}
         \caption{Host.}
         \label{fig:image_process_lock_count}
     \end{subfigure}
     \hfill
     \begin{subfigure}[b]{0.24\textwidth}
         \centering
         \includegraphics[width=\columnwidth]{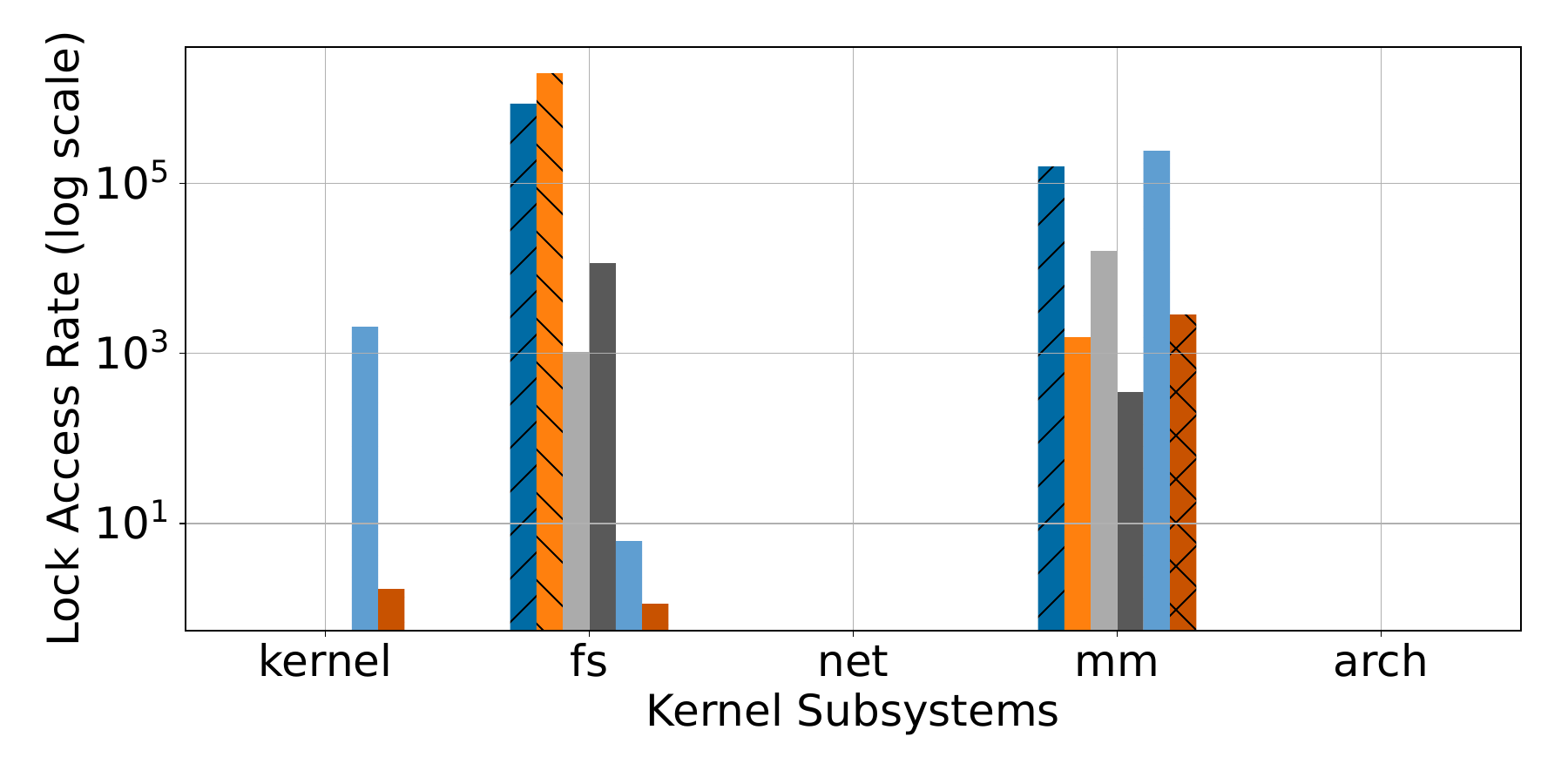}
          {\vspace{-0.25in}}
         \caption{runc.}
         \label{fig:video_process_count}
     \end{subfigure}
      \hfill
     \begin{subfigure}[b]{0.24\textwidth}
         \centering
         \includegraphics[width=\columnwidth]{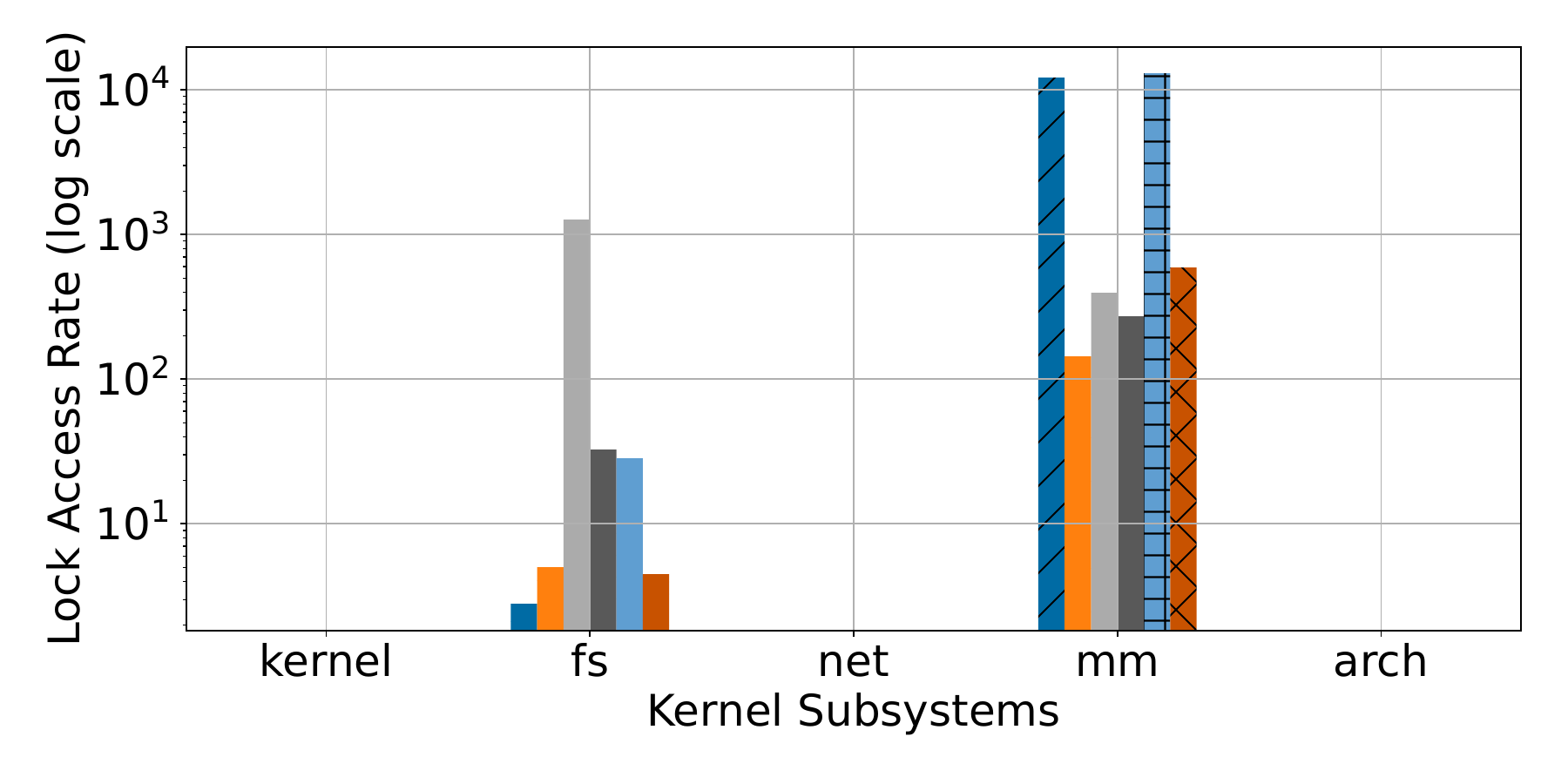}
          {\vspace{-0.25in}}
         \caption{runsc.}
         \label{fig:lr_lock_count}
     \end{subfigure}
      \hfill
      \begin{subfigure}[b]{0.24\textwidth}
         \centering
         \includegraphics[width=\columnwidth]{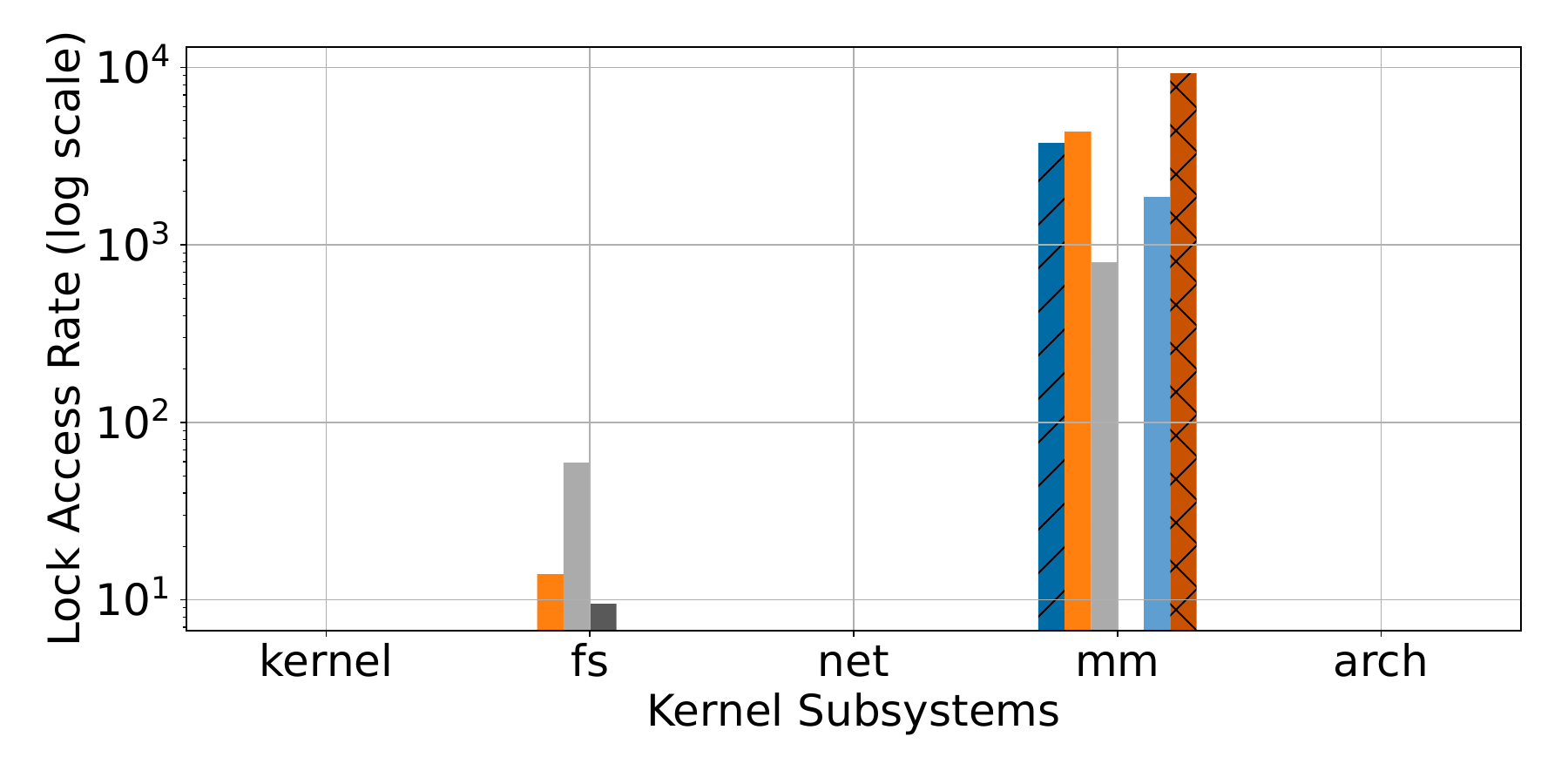}
          {\vspace{-0.25in}}
         \caption{fc.}
         \label{fig:fd_lock_count}
     \end{subfigure}
      \includegraphics[width=0.8\textwidth]{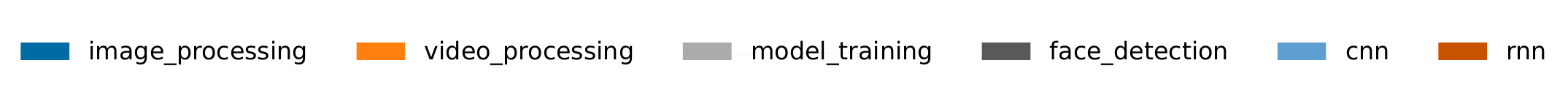}
          {\vspace{-0.18in}}
     \caption{Serverless workloads shared locks access rate across kernel subsystems. \textit{mm} and \textit{fs} are highly accessed across platforms.}
    \label{fig:kernel_funcbench_common}
\end{figure*}

\myparab{Shared lock count and access rate.} As observed in Table~\ref{tab:lock_count_rate}, host has the highest number of shared lock counts, followed by runc, runsc, and fc for all workloads. The shared lock rate for fc and runsc is lower compared to host and runc for most workloads (fc has the highest rate for RNN, and runsc has higher rate than host for CNN), suggesting that these more isolated environments experience less frequent interference on shared locks for these workloads.

Host and runc have the same highest accessed shared locks (\texttt{j\_state\_lock} and \texttt{journal\_s}) for image processing and video processing. RNN shows the lowest access rate for host, while runc has the lowest rate for face detection.

For most workloads, \texttt{zone->lock} for memory pages is one of the highly accessed locks by fc and runsc. \texttt{j\_state\_lock} is also acquired by runsc and fc for model training. As expected, we do not observe any networking lock. This indicates that journaling and memory management are potential sources of system-level interference for these serverless workloads, even in stronger isolation platforms like runsc and fc. 

\myparab{Kernel Subsystem.} As shown in Figure~\ref{fig:kernel_funcbench_common}, most shared locks across platforms are concentrated in the \textit{mm}, and \textit{fs} subsystems. 
\textit{fs} usage dominates on the host, particularly for model training and video processing workloads. There is considerable diversity in subsystem usage, suggesting deep reliance on kernel services. CNN and RNN exhibit the lowest usage of \textit{fs} locks. runc shows a similar trend with sharing across multiple subsystems, with the highest activity in \textit{fs} and \textit{mm}. In contrast, runsc acquires fewer shared locks, with activity concentrated in \textit{fs} and \textit{mm}. fc exhibits the least sharing, with shared locks confined to either \textit{fs} or \textit{mm}.

There is a wide variation in how all these workloads use kernel objects based on the platform they are using and the amount of stress they put on different subsystems, indicating the impact of design choices in kernel object usage. The same workload behaves differently in terms of kernel usages, signifying the importance of the \textit{right} platform choice for a given workload. 

Moreover, we note that serverless workloads are typically stateless, so frequent writing of file data is surprising may be due to using legacy code in a serverless environment. This also suggests that file systems for serverless workloads do not need crash consistency, so journaling could be disabled.


\subsubsection{Cloud Workloads}
\begin{table}[]
\small
\begin{tabular}{|l|l|}
\hline
\textbf{Benchmarks} & \textbf{\begin{tabular}[c]{@{}l@{}}Libraries/\\  SW\end{tabular}} \\ \hline
Graph Analytics & Apache Spark \\ \hline
Data Caching & Memcached \\ \hline
\begin{tabular}[c]{@{}l@{}}Feedsim\\ (Object Aggregation, \\ Ranking/Inference)\end{tabular} & \begin{tabular}[c]{@{}l@{}}Oldisim Library, ZLIB, Boost, OpenSSL, \\ BZIP2, LZ4, Snappy, libevent, jemalloc, \\ lzma, libsodium, rsocket, fmt, FBThrift, \\ Folly, wangle, fizz\end{tabular} \\ \hline
\begin{tabular}[c]{@{}l@{}}VideoTranscode\\ (Video Processing)\end{tabular} & ffmpeg, svt-av1, libaom \\ \hline
\end{tabular}
\caption{Cloud Workloads from DCPerf and CloudSuite.}
{\vspace{-0.23in}}
\label{ref:cloud_workloads_desc}

\end{table}

\begin{figure*}[t]
\centering
     \begin{subfigure}[b]{0.33\textwidth}
         \centering
         \includegraphics[width=\columnwidth]{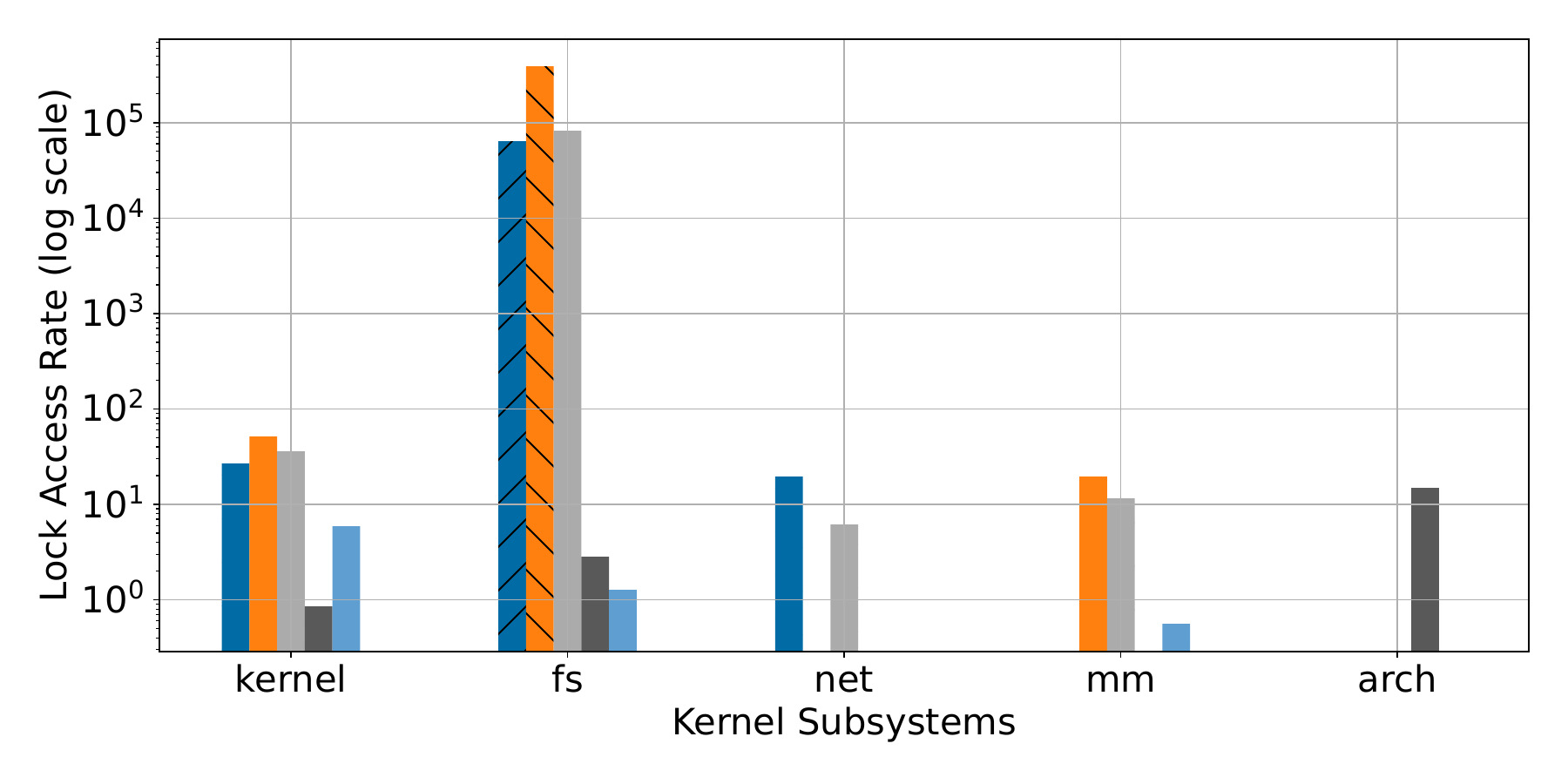}
         \caption{runc.}
         \label{fig:video_process_count}
     \end{subfigure}
      \hfill
     \begin{subfigure}[b]{0.33\textwidth}
         \centering
         \includegraphics[width=\columnwidth]{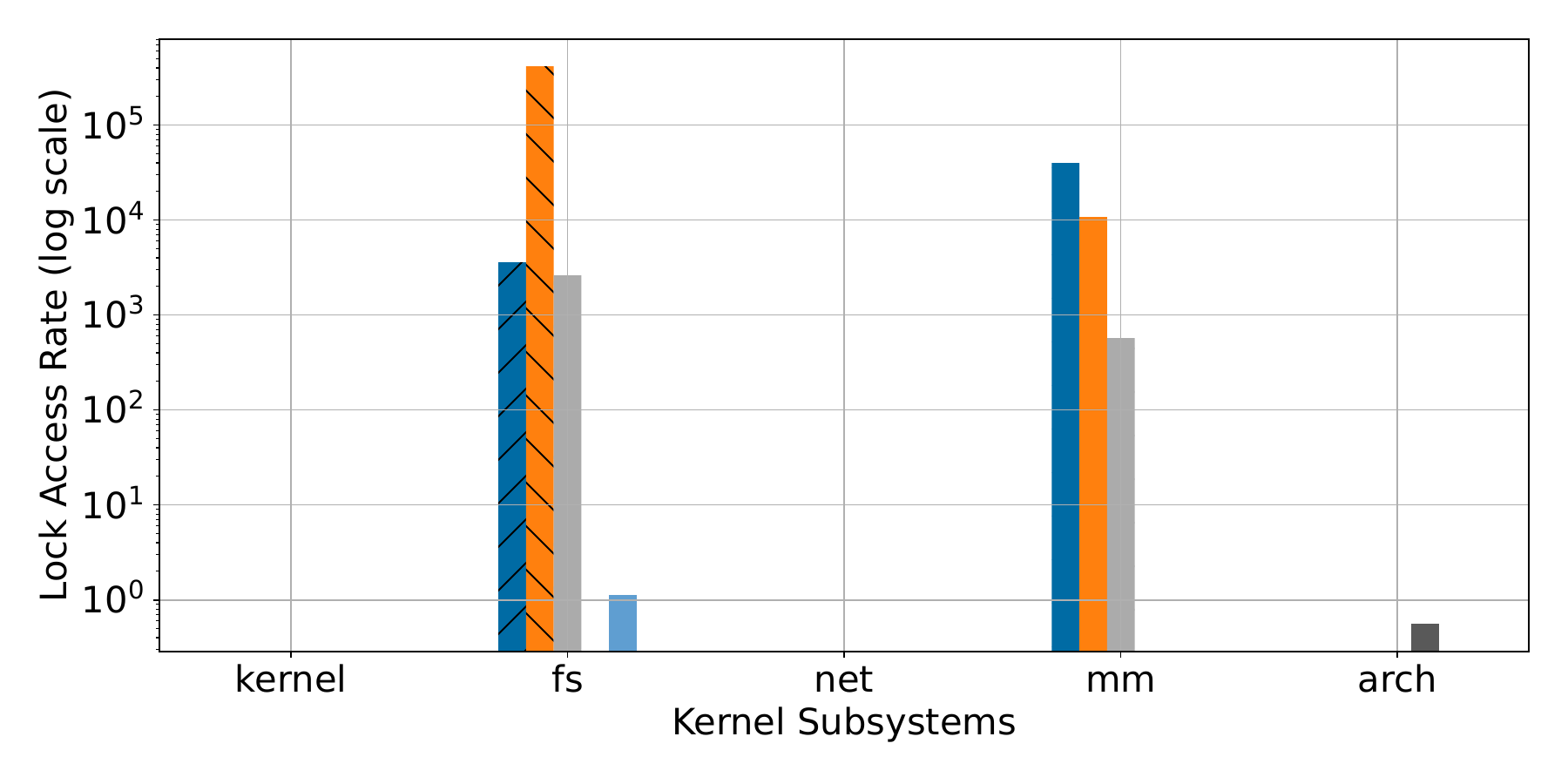}
         \caption{runsc.}
         \label{fig:lr_lock_count}
     \end{subfigure}
      \hfill
      \begin{subfigure}[b]{0.33\textwidth}
         \centering
         \includegraphics[width=\columnwidth]{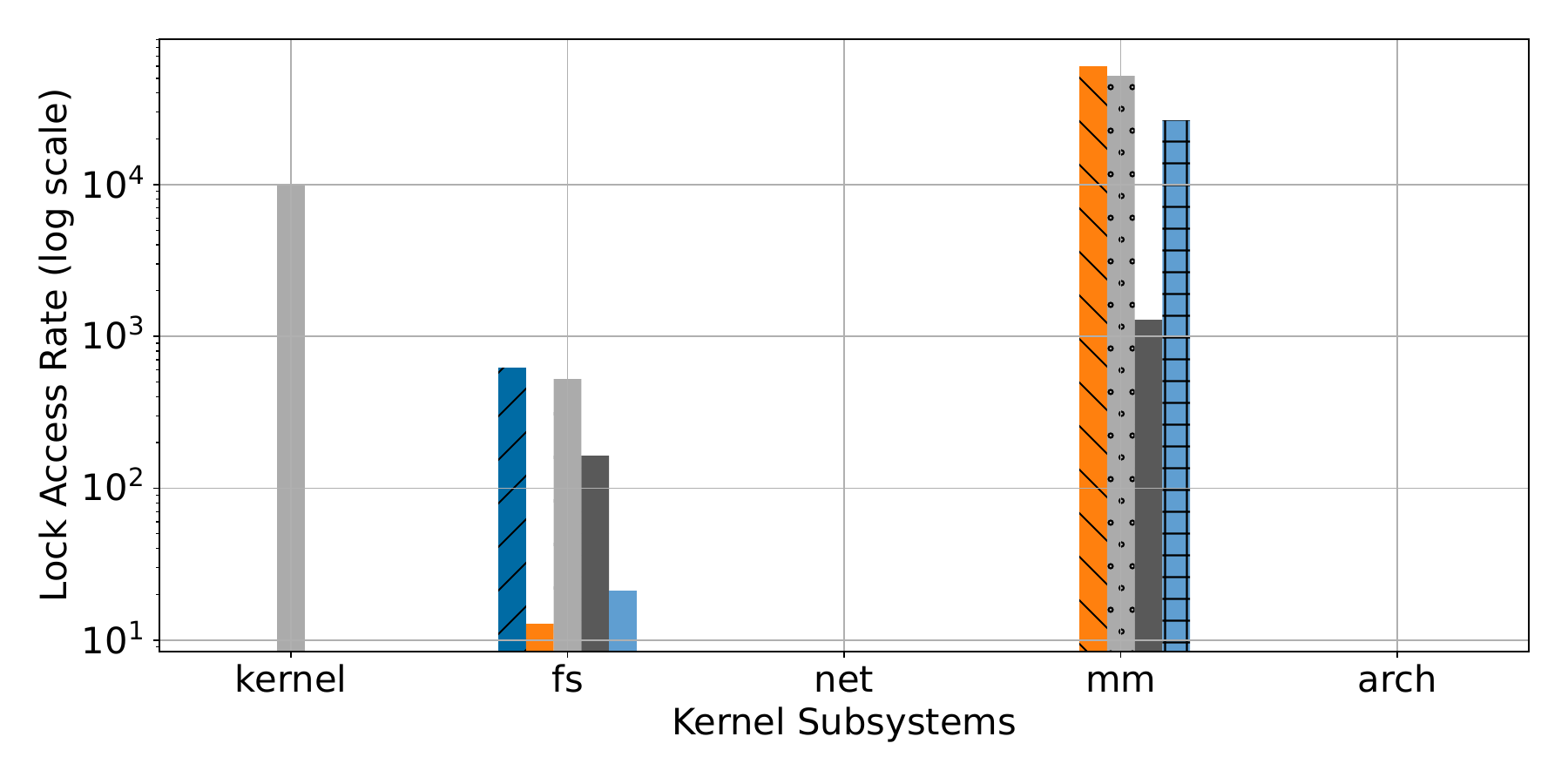}
         \caption{fc.}
         \label{fig:fd_lock_count}
     \end{subfigure}
     \includegraphics[width=0.85\textwidth]{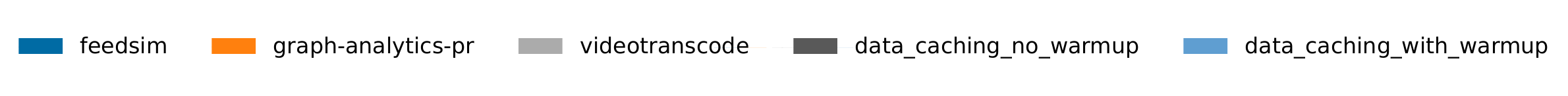}
      {\vspace{-0.18in}}
     \caption{Cloud workloads shared lock access rate across kernel subsystems. Most workloads show high usage in \textit{mm} and \textit{fs} across platforms.}
    \label{fig:kernel_common_cloud}
\end{figure*}

We use \toolname to trace four cloud workloads across different application domains, as shown in Table~\ref{ref:cloud_workloads_desc}. 
We run two concurrent workload instances, each with 24GB of memory and pinned to four separate cores on the same socket.

We run all workloads in standalone mode \ie all the services and components (like client and server) of the workload run in a single instance of the platform. While Feedsim and VideoTranscode do not have a separate server/client setup, we modified the other two workloads to run in standalone mode. Because the workload depends on virtual networking, we cannot run them directly on the host and only report results for runc, runsc, and fc.

\myparab{Graph Analytics} uses the Spark framework to perform graph analytics on large-scale datasets.
We run the PageRank algorithm for three iterations on a Twitter dataset with Spark's driver and executor memory set to 8GB each.

\myparab{Data Caching} runs a Memcached server simulating a Twitter data caching workload using a 10GB dataset.  We trace this setup in two modes: (a) with server warmup, and (b) without server warmup (tracing is skipped for warmup phase). 

\myparab{Feedsim}~\cite{ref:feedsim} represents the aggregation and ranking workloads in recommendation systems. It searches for the maximum QPS that the system can achieve while keeping p95 latency to be no greater than 500ms. 

\myparab{VideoTranscodeBench}~\cite{ref:video_transcode} is based on ffmpeg representing video encoding workloads. It can apply different encoders and videos, and run them at various encoding levels. 

\myparab{Shared lock count and access rate.}  runc and runsc have the highest shared locks for VideoTranscode. For all other workloads, runc has the highest number of shared locks, while fc has the least, except for data caching (in both modes). 

runc has the highest rate for feedsim and videotranscode, with most locks belonging to the filesystem. The top two locks are  \texttt{root->kernfs\_rwsem} (protects \texttt{kernfs\_root} responsible for maintaining the root of the kernel virtual file system) and \texttt{journal->j\_state\_lock}. The journal lock remains the top accessed for graph analytics as well. Notably, it does not exhibit any significant access to data caching. 

Similar to runc, runsc also grabs mostly filesystem (\texttt{journal->j\_state\_lock}, \texttt{bgl->locks[i].lock} and \texttt{root->kernfs\_rwsem} with a higher rate) locks and \texttt{zone->lock} for most workloads. Data caching has minimal sharing for runsc. 

\texttt{lruvec->lru\_lock} is the highest source of interference for fc based on rate for all workloads but videotranscode. It uses a handful of filesystem locks for videotranscode.

These results further indicate variation in kernel object accesses by different workloads based on the platform they use, varying their isolation level via system structures. 

Although there are shared lock accesses by these workloads, but the rate with which they are accessed is lower compared to microbenchmarks and serverless for most, indicating low resource pressure.

\myparab{Kernel Subsystem.} We make a similar observation compared to serverless workloads: filesystems, and memory management subsystems remain a significant source of system-level interference.

\subsection{Lock Interference Summary}
These results demonstrate a high variance in shared lock access, predicting variation in the interference of co-located workloads. They also show the data structure likely to cause interference, namely the page allocator and reclamation mechanism, and the file system journal. While much of the file system uses fine-grained locking, such as the inode and dentry caches, the journal is a single point of contention.

We also note that both coarse and fine-grained sharing can contribute to object-level interference, which can lead to performance overhead during high accesses under load.



\section{Performance Interference}
\label{sec:perf_interference}
The preceding section looked at the level of shared locking across workloads. In this section we evaluate the performance interference through system resources for the same workloads, showing that the level of shared locking relates to the amount of interference.

\subsection{Microbenchmarks}
We run the same microbenchmarks stressing different kernel subsystems from Section~\ref{sec:locks_interference} to understand how sharing different kernel objects impacts the performance of co-running workloads. As before, we do not evaluate CPU-only workloads.

\subsubsection{Memory}
For this benchmark, we run one worker and seven trashers (Section~\ref{eval:perf-interference-trasher}). The worker runs for 80 minutes, and the first trasher starts after ten minutes, and every ten minutes, an additional trasher starts. 
We run the benchmark three times and report the average performance. Between each run, we reset the environment.
 
\begin{figure*}[t]
\centering
     \begin{subfigure}[b]{0.33\textwidth}
         \centering
         \includegraphics[width=\columnwidth]{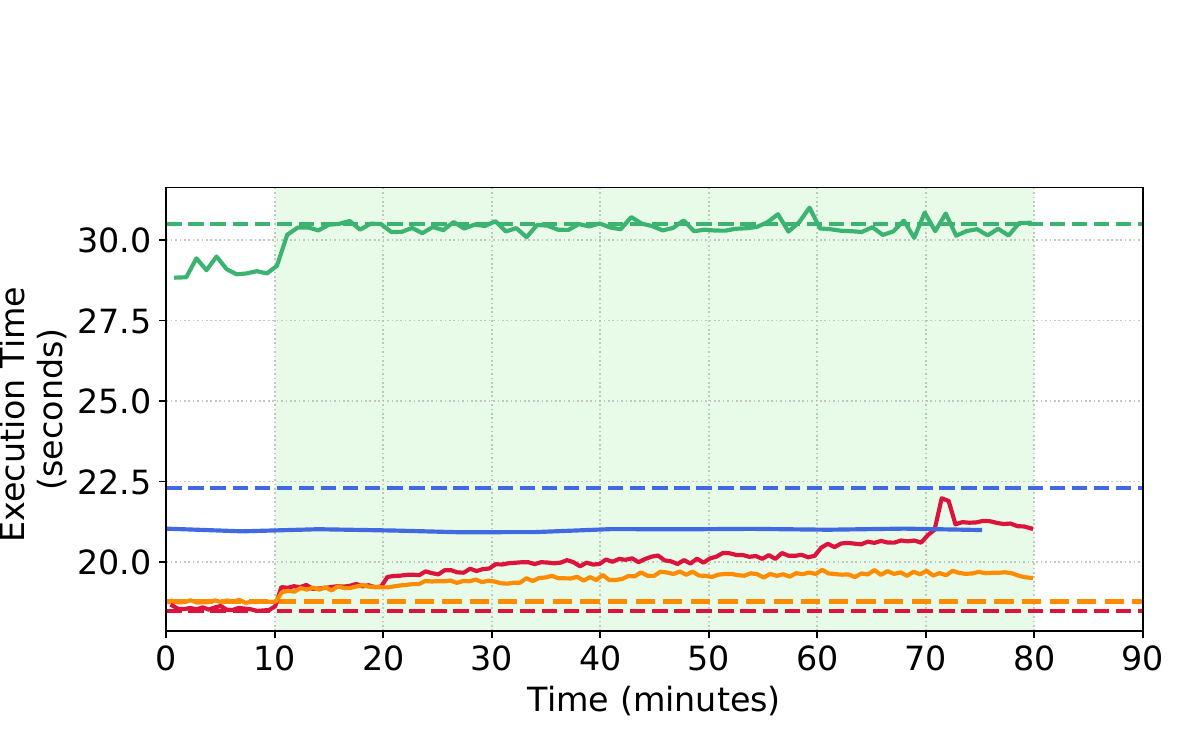}
         {\vspace{-0.25in}}
         \caption{Allocation time (8KB).}
         \label{fig:alloc_8KB}
     \end{subfigure}
     \hfill
     \begin{subfigure}[b]{0.33\textwidth}
         \centering
         \includegraphics[width=\columnwidth]{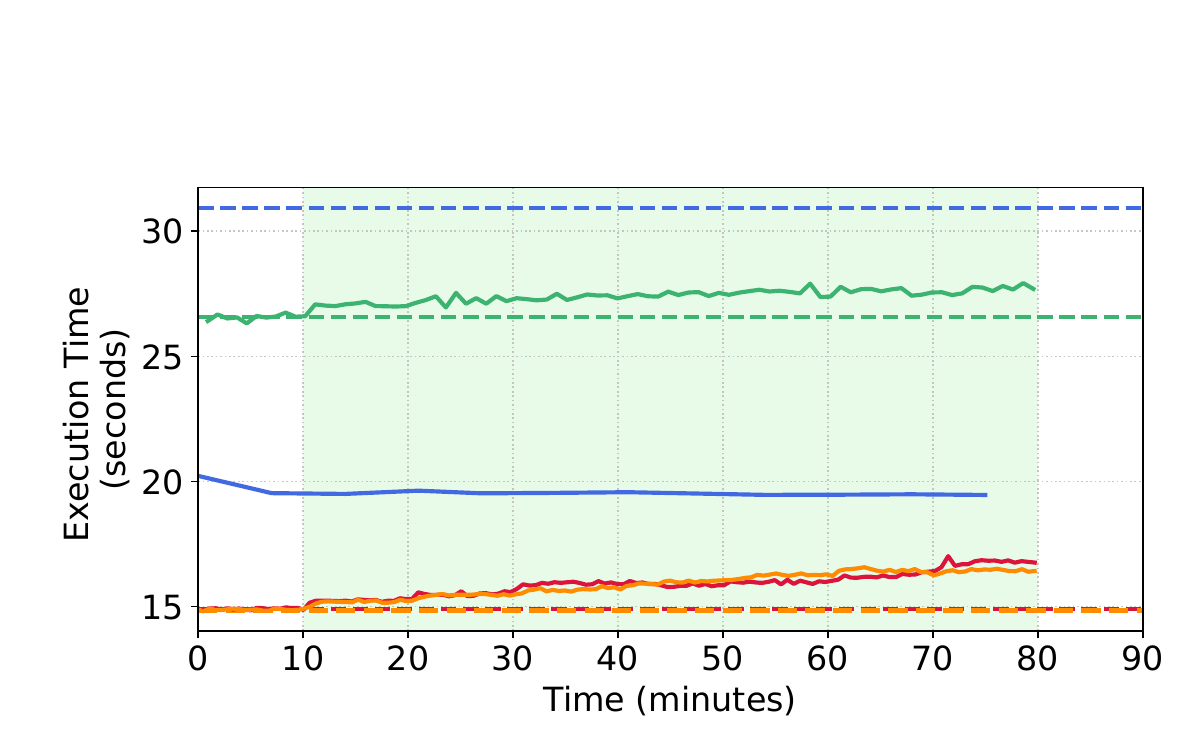}
         {\vspace{-0.25in}}
         \caption{Touch time (8KB).}
         \label{fig:touch_8KB}
     \end{subfigure}
      \hfill
     \begin{subfigure}[b]{0.33\textwidth}
         \centering
         \includegraphics[width=\columnwidth]{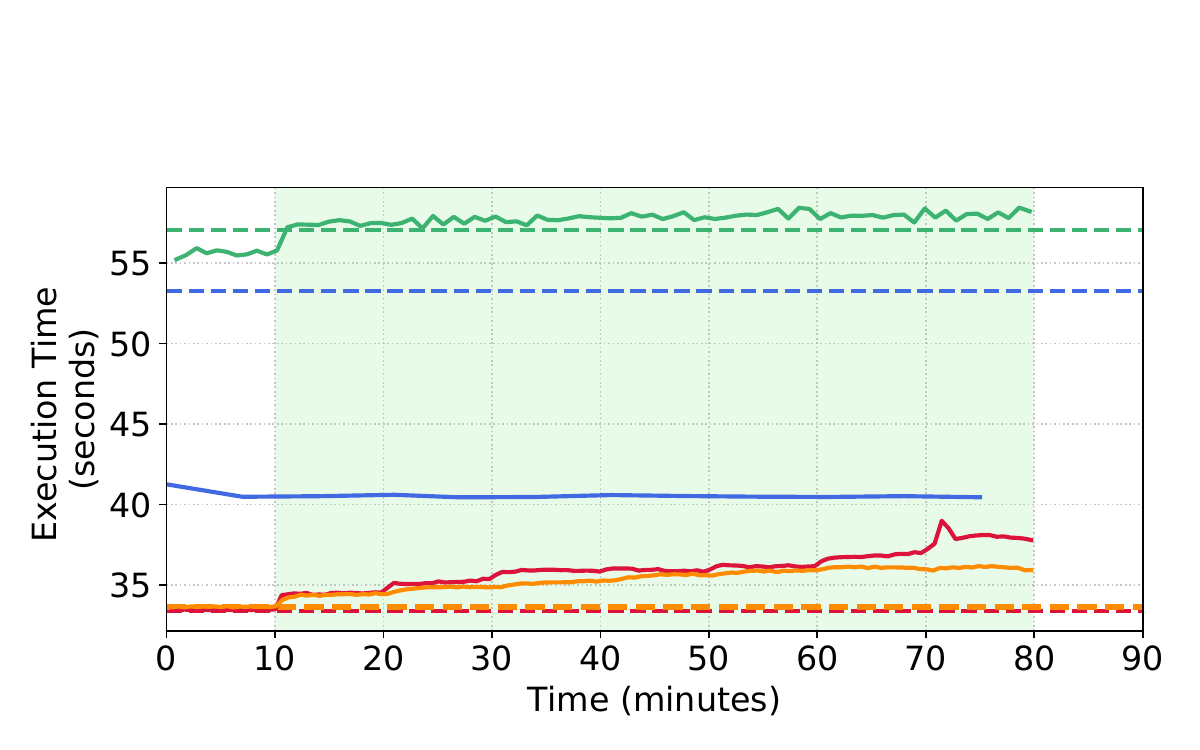}
         {\vspace{-0.25in}}
         \caption{Total (allocation + touch) time (8KB).}
         \label{fig:total_8KB}
     \end{subfigure}    

{\vspace{-0.05in}}
    \hfill
     \begin{subfigure}[b]{0.33\textwidth}
         \centering
         \includegraphics[width=\columnwidth]{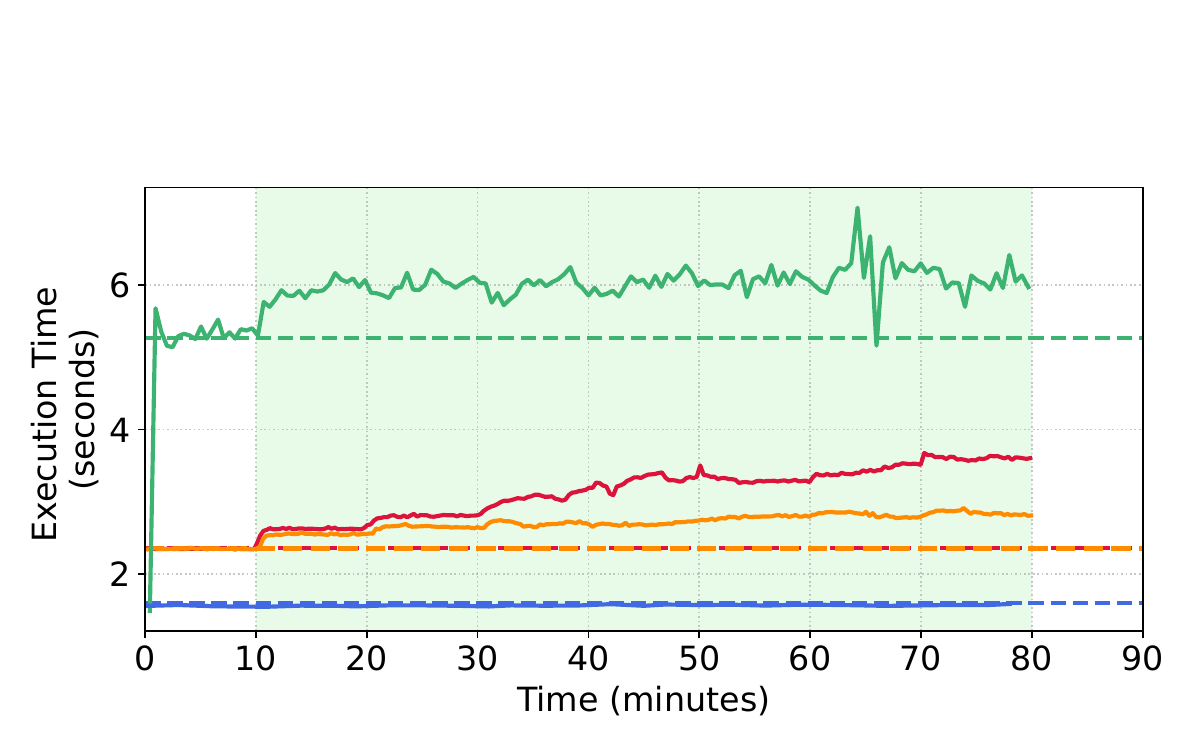}
         {\vspace{-0.25in}}
         \caption{Allocation time (1MB).}
         \label{fig:alloc_1MB}
     \end{subfigure}
     \hfill
     \begin{subfigure}[b]{0.33\textwidth}
         \centering
         \includegraphics[width=\columnwidth]{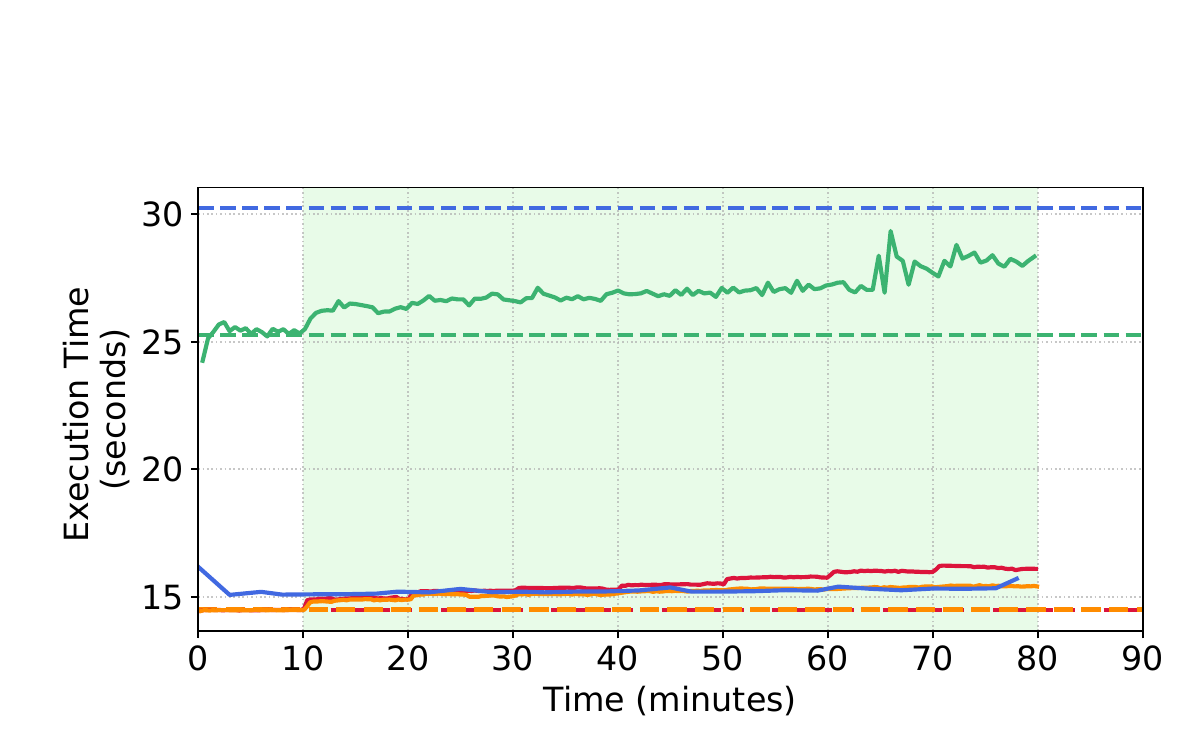}
         {\vspace{-0.25in}}
         \caption{Touch time (1MB).}
         \label{fig:touch_1MB}
     \end{subfigure}
      \hfill
     \begin{subfigure}[b]{0.33\textwidth}
         \centering
         \includegraphics[width=\columnwidth]{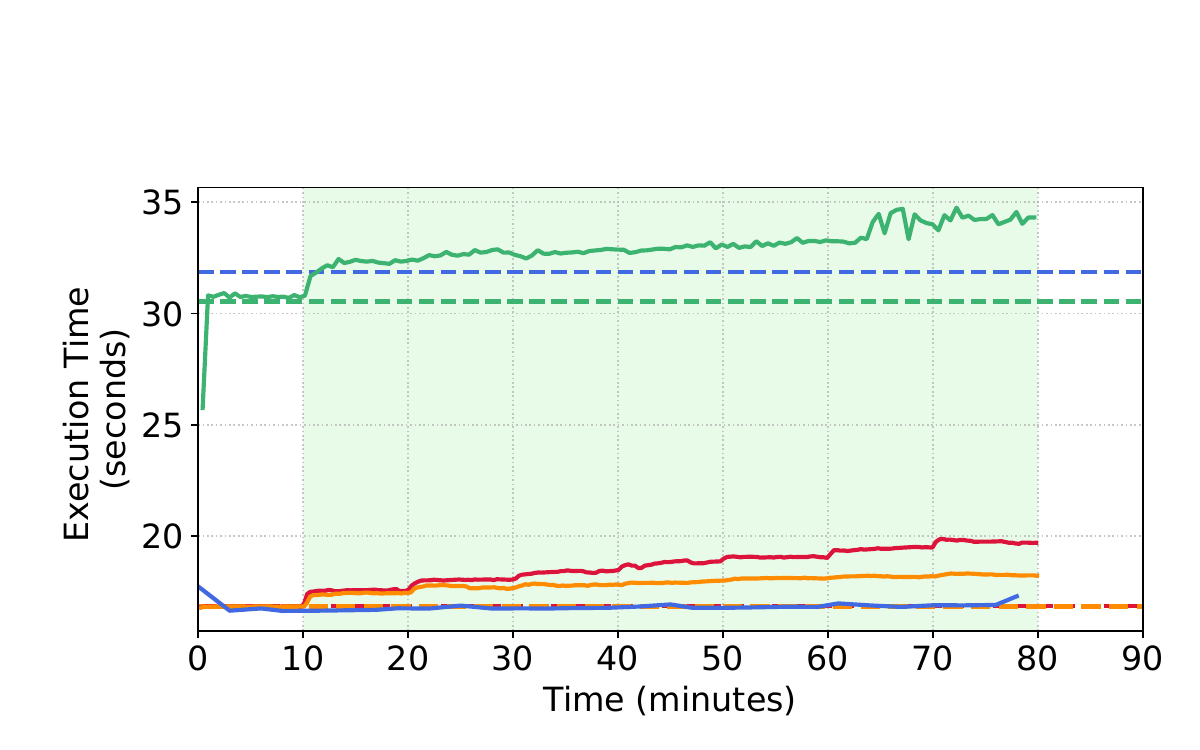}
         {\vspace{-0.25in}}
         \caption{Total (allocation + touch) time (1MB).}
         \label{fig:total_1MB}
     \end{subfigure}
     \includegraphics[width=0.95\textwidth]{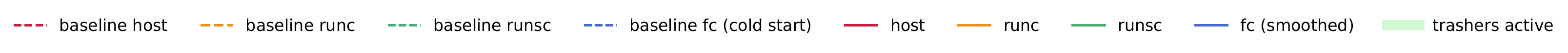}
     {\vspace{-0.18in}}
     \caption{Memory stress tests for 8KB (~\ref{fig:alloc_8KB}-~\ref{fig:total_8KB}) and 1MB (~\ref{fig:alloc_1MB}-~\ref{fig:total_1MB}) allocation sizes for a total memory of 16GB. All but fc show performance degradation over time under load. Note that y-axes do not go to zero. Lower numbers are better.}
    \label{fig:mem_perf}
    {\vspace{-0.18in}}
\end{figure*}

We observe the effects of object-level interference on host, runc, and runsc for both allocation sizes in Figure~\ref{fig:mem_perf}. gVisor has two levels of virtual-to-physical page mappings~\cite{ref:runsc_mm}, one from the application to Sentry and the other from Sentry to the host. This leads to a lower baseline performance compared to \textit{host} and runc. As noted, runsc and host acquire the global allocator lock \texttt{zone->lock} with a high rate, which becomes a point of high interference for both under memory load, as indicated by their performance degradation. 
As the load increases in the system, we can see the impact of such sharing on runc from its degraded performance.     


Firecracker is not impacted by other co-running microVMs over time as the total mmap size remains consistent across iterations. After the initial boot, where it maps the necessary pages into guest memory, fc does not need to make additional \texttt{mmap} calls to the host to allocate more physical memory for subsequent allocations. The guest OS within the microVM allocates and deallocates within the already mapped region for subsequent allocations. However, it does has have a higher execution time after startup, which comes from a high rate of access to \texttt{zone->lock} and \texttt{lruvec->lru\_lock} due to initial page allocations.


\subsubsection{Filesystem Metadata Operations}
\begin{figure*}[t]
\centering
     \begin{subfigure}[b]{0.33\textwidth}
         \centering
         \includegraphics[width=\columnwidth]{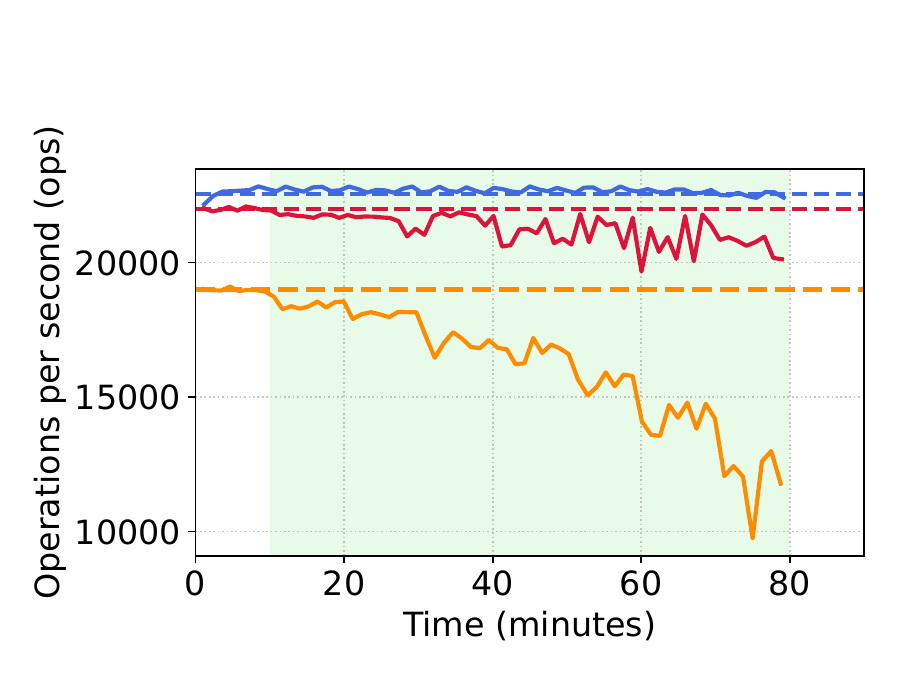}
         {\vspace{-0.25in}}
         \caption{List directory.}
         \label{fig:filebench_list}
     \end{subfigure}
     \hfill
     \begin{subfigure}[b]{0.33\textwidth}
         \centering
         \includegraphics[width=\columnwidth]{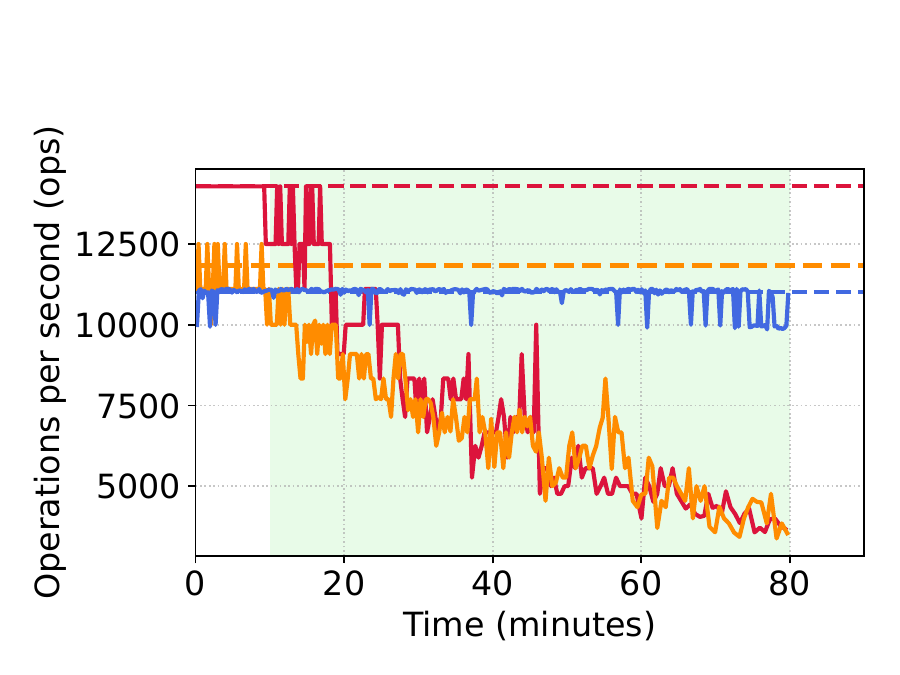}
         {\vspace{-0.25in}}
         \caption{Create and close files.}
         \label{fig:filebench_create}
     \end{subfigure}
      \hfill
     \begin{subfigure}[b]{0.33\textwidth}
         \centering
         \includegraphics[width=\columnwidth]{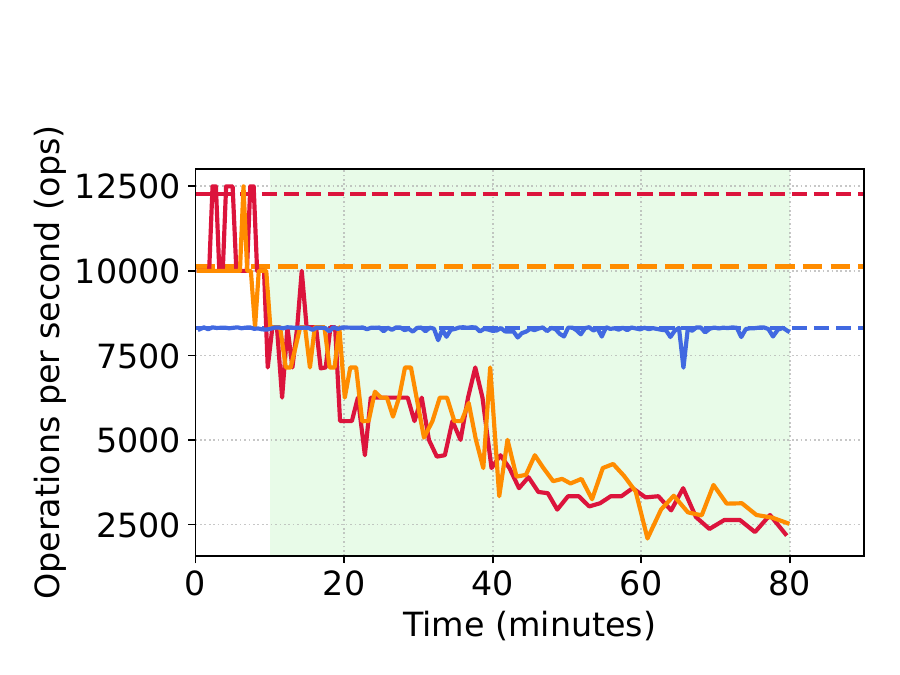}
         {\vspace{-0.25in}}
         \caption{Delete files.}
         \label{fig:filebench_delete}
     \end{subfigure}
     \includegraphics[width=0.9\textwidth]{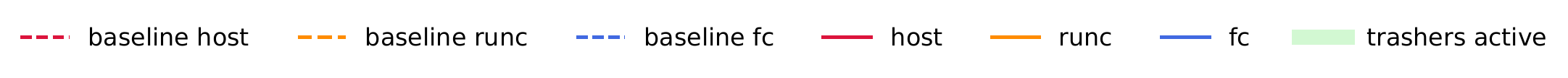}
     {\vspace{-0.18in}}
     \caption{Filebench stress tests for filesystem metadata operations. Both, host and runc show significant performance degradation for create/close and delete operations. fc performs uniformly under load. Note that y-axes do not go to zero. Higher numbers are better.}
     {\vspace{-0.15in}}

    \label{fig:filebench_perf}
\end{figure*}

For all metadata benchmarks, we observe performance degradation for host and runc with fc having the lowest baseline for create and delete metadata operations as shown in Figure~\ref{fig:filebench_perf}. We calculate the baseline by running a single instance for each for ten minutes and averaging over the iterations completed during that time. fc has stable performance for these operations under stress, with some minor degradation at times. Filesystem locks acquired during metadata operations showed a very high access rate from our lock usage analysis, and we can see the impact of those locks' interference (journaling locks in particular) on the degraded performance on all three platforms. Even incidentally shared locks like \texttt{inode\_hash\_lock} acquired by runc for metadata operations contribute to significant object interference under pressure, resulting in performance degradation. 

host and runc are impacted more as they share more of the host kernel's filesystem internal state compared to fc, where isolation at the microVM level limits such interference between instances.


\subsection{Serverless Workloads}
\label{sec:serverless_perf}

\begin{figure*}[t]
\centering
     \begin{subfigure}[b]{0.33\textwidth}
         \centering
         \includegraphics[width=\columnwidth]{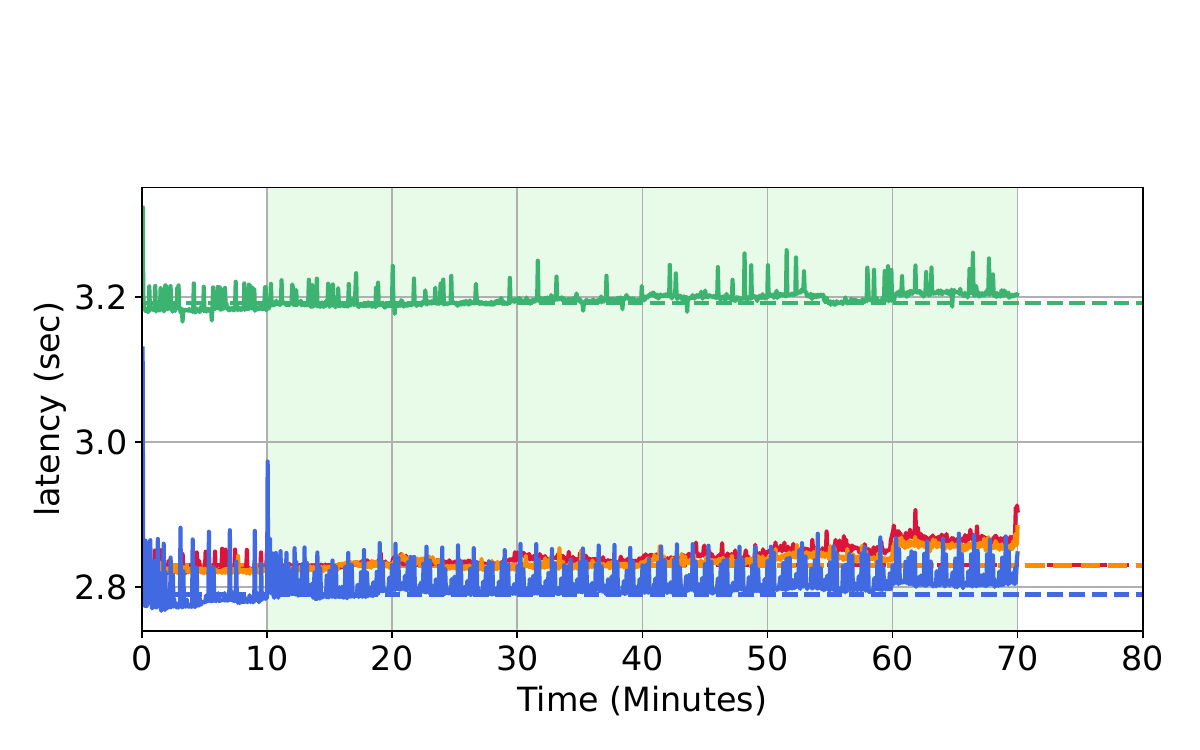}
         \caption{Image Processing.}
         \label{fig:image_process_perf}
     \end{subfigure}
     \hfill
     \begin{subfigure}[b]{0.33\textwidth}
         \centering
         \includegraphics[width=\columnwidth]{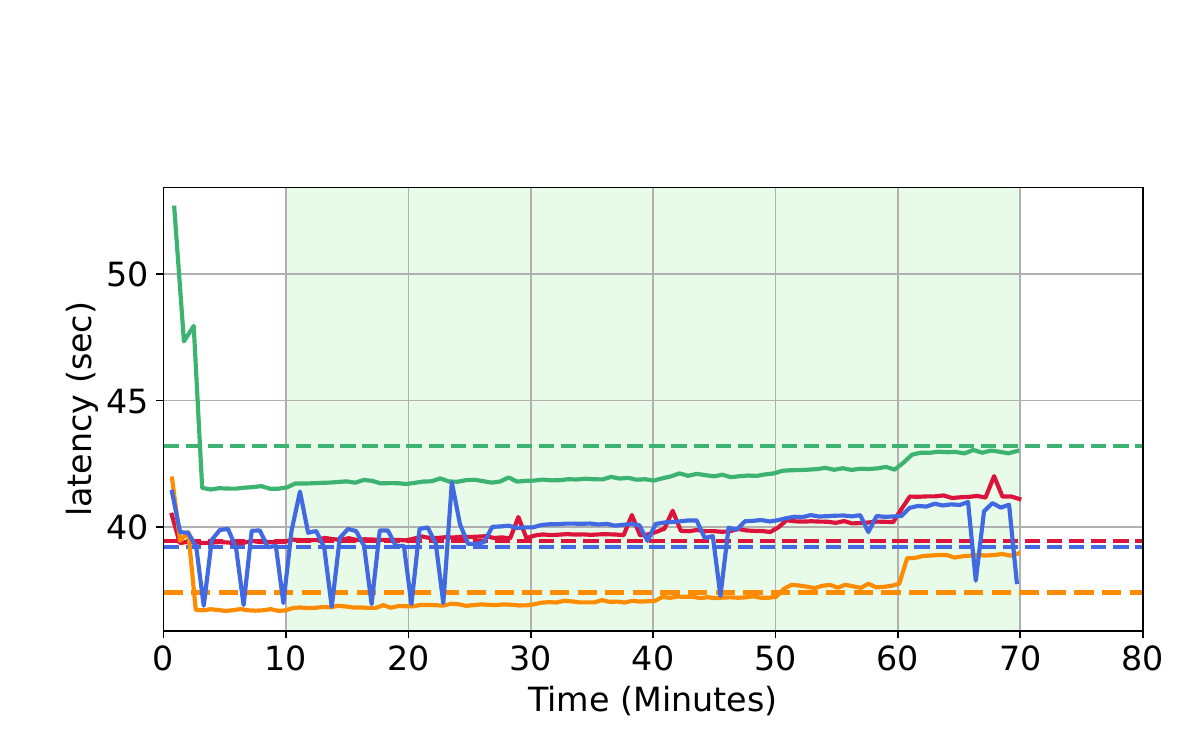}
         \caption{Video Processing.}
         \label{fig:video_process_perf}
     \end{subfigure}
      \hfill
     \begin{subfigure}[b]{0.33\textwidth}
         \centering
         \includegraphics[width=\columnwidth]{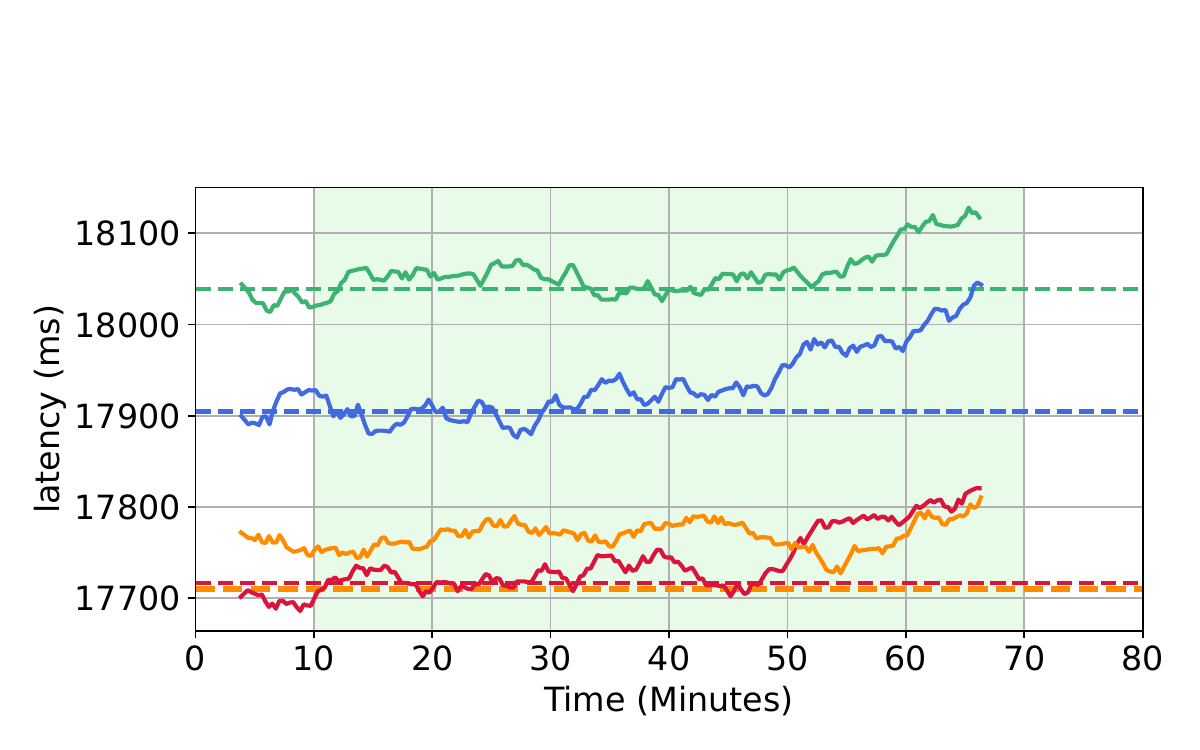}
         \caption{Model Training.}
         \label{fig:lr_perf}
     \end{subfigure}
     {\vspace{-0.19in}}

      \hfill
      \begin{subfigure}[b]{0.33\textwidth}
         \centering
         \includegraphics[width=\columnwidth]{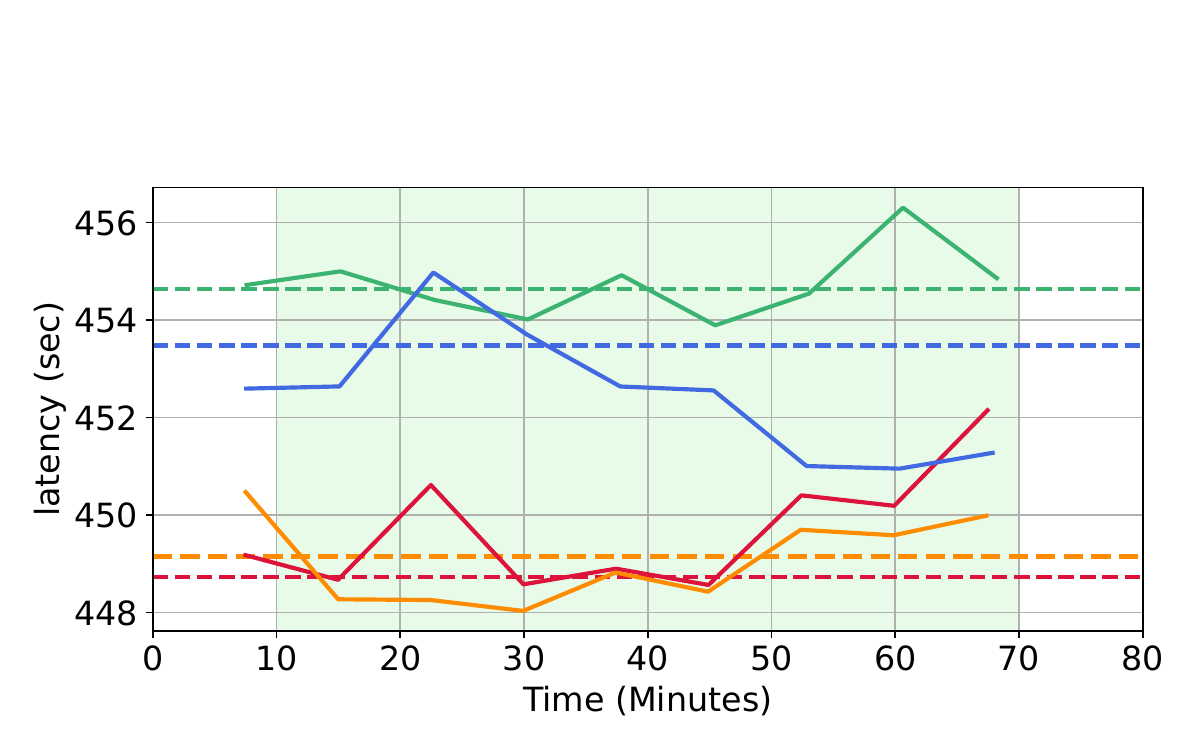}
         \caption{Face Detection.}
         \label{fig:fd_perf}
     \end{subfigure}
     \hfill
     \begin{subfigure}[b]{0.33\textwidth}
         \centering
         \includegraphics[width=\columnwidth]{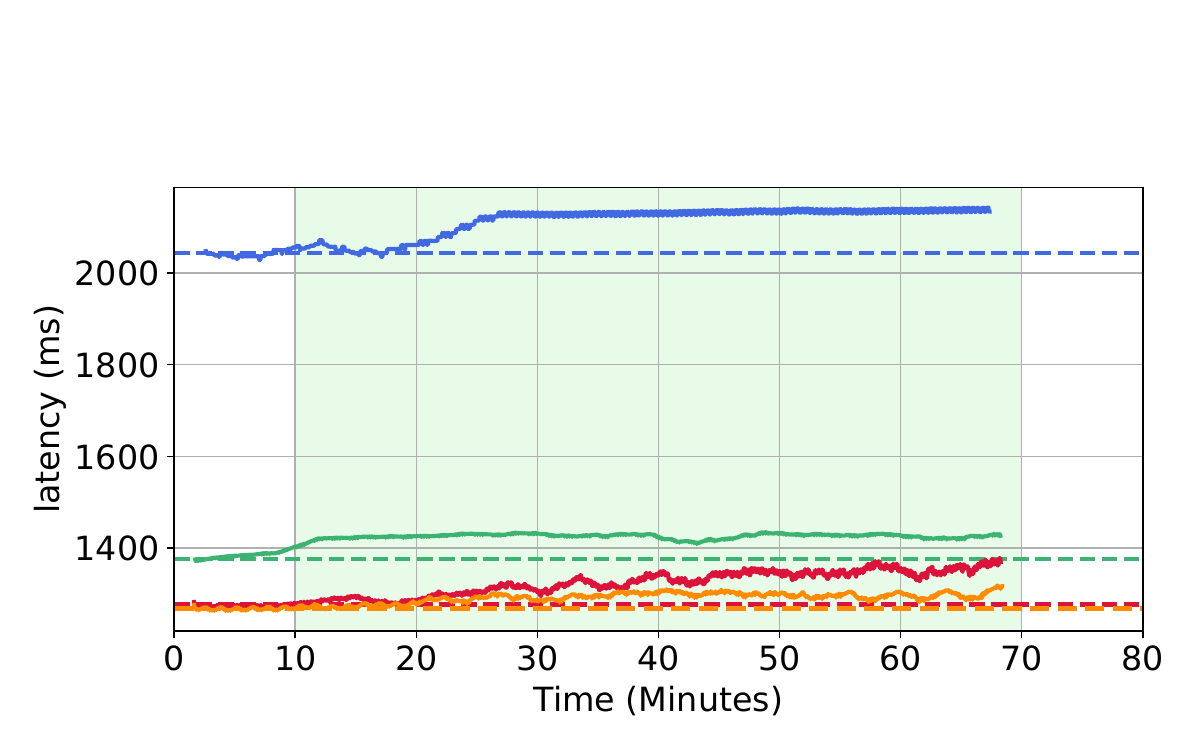}
         \caption{CNN.}
         \label{fig:cnn_perf}
     \end{subfigure}
      \hfill
     \begin{subfigure}[b]{0.33\textwidth}
         \centering
         \includegraphics[width=\columnwidth]{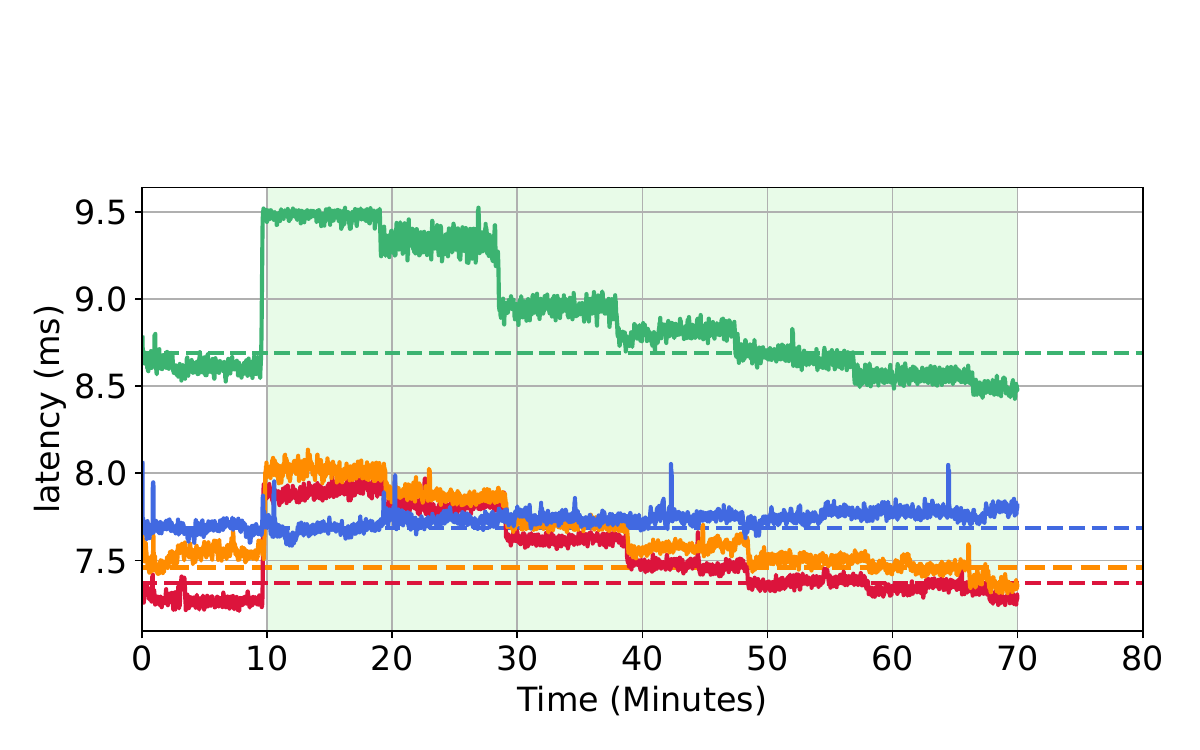}
         \caption{rnn.}
         \label{fig:rnn_perf}
     \end{subfigure}
     \includegraphics[width=0.95\textwidth]{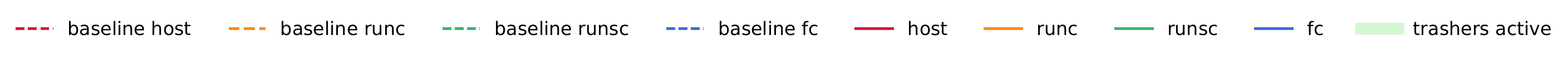}
     {\vspace{-0.18in}}
     \caption{FunctionBench stress tests. All platforms show some performance degradation under load for most workloads. Note that the Y-axes do not go to zero. Lower numbers are better.}
    \label{fig:funcbench}
\end{figure*}

We show the performance of the worker over time for all serverless workloads in Figure~\ref{fig:funcbench}. Similar to the microbenchmarks, we calculate the baseline by running a single instance with no co-runner interference and averaging the results across multiple runs, and compare against an instance with increasing numbers of trashers.

We observe performance degradation across several workloads on all platforms. Our lock usage analysis revealed that these workloads exhibit a wide range of lock access patterns, with high acquisition rates across most workloads, particularly in the filesystem and memory subsystems.

These workloads involve complex sharing patterns, ranging from fine-grained locks such as \texttt{bgl->locks[i].lock} on ext4 blockgroups, to incidentally shared locks like \texttt{inode\_hash\_lock}, and several global locks (\texttt{zone->lock}, several journaling locks) acquired along multiple kernel code paths. These overlapping access patterns lead to numerous points of object-level interference within shared kernel subsystems.

The result is compounded performance degradation, as seen in the  Figure~\ref{fig:funcbench}. Notably, no single platform consistently outperforms the others across all workloads in terms of providing isolation. The degree to which each platform isolates kernel objects, and thus avoids lock-level interference, determines which workloads benefit more from a given isolation platform.

\subsection{Cloud Workloads}
We run one (baseline) and two instances to measure performance interference for cloud workloads rather than using trashers. We observe minimal performance degradation (less than 1\%) across these workloads. This is likely due to limited resource pressure, as only two instances are running. Our lock usage analysis supports this observation, showing relatively low lock access rates for these workloads, insufficient to cause significant interference under this low load.

\subsection{Performance Interference Summary}
Our performance evaluation reveals that object-level sharing can significantly impact performance as access rates to shared kernel objects increase. This has important implications for workload scheduling in multi-tenant environments: co-located workloads that frequently access the same shared kernel objects can interfere with one another, with the extent of interference varying based on the isolation platform in use.
\section{Interference Analysis Use-Cases}
\label{sec:use_cases}
We envision multiple use cases for our analysis of kernel lock interference. 

\myparab{Quantify isolation via a metric.} We can use the lock interference data to formulate a metric that measures isolation by measuring the amount of sharing and the impact it has on interference. This metric can be used in scheduling workloads in the cloud to minimize interference in the cloud. \eg interfering workloads mostly reading shared objects are less likely to interfere and can be safely scheduled together. 

\myparab{Identify the type of sharing.} By identifying objects that are heavily shared vs those that are not can help us understand the nature of sharing. Sharing in the kernel varies across different subsystems. Sharing is necessary to ensure consistent access to globally visible resources, such as file data.  In other cases, sharing is used for convenience to create simpler data structures. \eg the futex table is shared by all processes, but each process only accesses its private locks in the table.

\myparab{Using the metric for design.} By identifying which data has to be shared and which need not, kernel developers can work to reduce the amount of interference possible between workloads, and eliminate or minimize things shared only for convenience.  This analysis can provide insights for application developers to optimize their implementations, minimizing data sharing where feasible. Additionally, it can guide isolation platform developers in enhancing their platforms by adding layers, similar to 'sentry' in gVisor, to segregate highly shared data structures.



\section{Related Work}
\label{sec:related}

\myparab{Kernel Locks} Multiple tools have been proposed to detect race conditions and deadlocks. Eraser~\cite{ref:eraser} proposes a novel lockset algorithm for dynamically detecting data races in multithreaded programs. ``Locksets'' consist of all held locks accessing shared variables. If the lockset for shared variable is empty then the variable is flagged as not being consistently protected.
LockDoc~\cite{ref:lockdoc} proposes a dynamic trace-based analysis of an instrumented kernel to infer locking rules of members of data structures to understand the implementation and to detect possible locking rule violations.

\myparab{Scalability and Commutativity} Min \etal~\cite{ref:scalability_fs} studied the scalability behavior of some popular file systems and identified some kernel objects as the source of scalability bottlenecks. Clements \etal \cite{ref:commutativity_rule_mit} introduced a scalable commutativity rule, suggesting that when interface operations are performed in any order without affecting the outcome, they can be implemented in a manner that allows for scalability. Multikernel ~\cite{ref:multikernel} proposes a new architecture for scaling multicore systems by avoiding any inter-core sharing.

\myparab{Interference Measurement} A recent work~\cite{ref:kit} detects inter-container functional interference by comparing the system call traces of a container across two different executions (running with and without another container). Some works have studied and measured performance interference for workloads by either characterizing workloads~\cite{ref:bolt, ref:parties} under different stress levels or by proposing an analytical model~\cite{ref:interferece_model}.

\myparab{Address Space Isolation} Address-space isolation (ASI)~\cite{ref:asi_1, ref:asi_2} is the technique of unmapping unnecessary kernel memory, making it inaccessible to the current running context. The implementation of ASI depends on marking memory as sensitive and non-sensitive; sensitive memory is unmapped, restricting the address space for system calls. 
A possible approach to identifying sensitive memory is to detect critical data structures and unmap their memory when not needed. Our analysis can serve as an initial step toward detecting these sensitive data structures.

\section{Conclusions}
\label{sec:conclusion}
In this paper, we introduced and evaluated a new approach to understanding and measuring system-level interference. By collecting and analyzing kernel-level lock activity, we identified heavily accessed kernel objects across a range of workloads and isolation platforms. Our results reveal significant variation in object access patterns, leading to differing performance implications across platforms. The analysis also highlights the file system and memory management subsystems as the most frequently accessed components across these platforms.



{
\bibliographystyle{plain}
\bibliography{paper}
}


\end{document}